\documentclass[11pt]{article}
\usepackage[spanish,es-tabla]{babel}
\usepackage{natbib}
\usepackage{comment}
\usepackage{float}
\usepackage[hidelinks]{hyperref}
\usepackage{mathrsfs}
\usepackage{enumitem}
\usepackage[font={small,it}]{caption}
\usepackage{amsmath,amsfonts,amsthm,amssymb}
\usepackage{bm,rotating,multirow,dsfont,graphicx}
\usepackage[usenames, dvipsnames]{color}
\usepackage{url}
\usepackage{multicol}
\usepackage{multirow}
\usepackage[T1]{fontenc}
\usepackage{flafter}
\usepackage{appendix}
\usepackage{subfigure}
\usepackage{xcolor}
\usepackage{ragged2e}
\usepackage{listings}
\usepackage{authblk}
\spanishdecimal{.}

\makeatletter
\def\hlinewd#1{%
	\noalign{\ifnum0=`}\fi\hrule \@height #1 %
	\futurelet\reserved@a\@xhline}
\makeatother
\usepackage[all,cmtip]{xy}

\def\frac#1#2{{{#1}\over{#2}}}

\def\@roman#1{\romannumeral #1}

\begin{document}

\renewcommand{\refname}{Referencias}

\def\spacingset#1{\renewcommand{\baselinestretch}{#1}\small\normalsize}\spacingset{1}

\title{Caracterización del discurso de posesión presidencial e identificación de comunidades políticas en Colombia: Aproximación empírica desde el análisis de redes sociales}

\author{
    Carolina Luque$^1$\footnote{Contacto: cluque2.d@universidadean.edu.co.} \\
    Isabella Agudelo$^2$\footnote{Contacto: isagudelog@unal.edu.co. Segundo autor conjunto.}\qquad Kevin Leal$^2$\footnote{Contacto: klealp@unal.edu.co. Segundo autor conjunto.} \\\vspace{0.3cm}
    Juan Sosa$^2$\footnote{Contacto: jcsosam@unal.edu.co.}}
\date{%
    $^1$Universidad Ean, Bogotá, Colombia\\\vspace{0.3cm}%
    $^2$Universidad Nacional de Colombia, Bogotá, Colombia\\%
    %\today
}

\maketitle

\begin{abstract} 
El análisis estadístico de redes sociales es una metodología predominante en la investigación en ciencias políticas. En este artículo proponemos implementar métodos de redes para caracterizar el discurso presidencial e identificar comunidades políticas en Colombia. Planteamos una aproximación empírica a la estructura discursiva de los jefes de Estado y a la configuración de relaciones de alianza y trabajo entre figuras prominentes de la política colombiana. Así, implementamos métodos de redes desde dos perspectivas, palabras y actores políticos. Concluimos sobre la relevancia de estadísticas de redes sociales para identificar términos frecuentes e importantes en la acción comunicativa de los gobernantes, y para examinar cohesión en los discursos. Finalmente, distinguimos actores notables en la consolidación de relaciones de trabajo y alianzas, y comunidades políticas.
\end{abstract}

\noindent
{\it Palabras clave:} Análisis del discurso, comunidades políticas, modelo de bloques estocásticos, redes sociales.

\spacingset{1.1} % DON'T change the spacing!

%%%%%%%%%%%%%%%%%%%%%%%%%%%%%%%%%%%%%%%%%%%%%%%%%%%%%%%%%%%%%%%%%%%%%%%%%%%%%%%%%%%%%%%%%%%%%%%%%%%%%%%%%%%%%%%%%%%%

\newpage

\section{Introducción}
El estudio de la información que emerge de la interconexión entre elementos autónomos en un sistema es fundamental para analizar diversos fenómenos. 
La estructura que forman los elementos (actores o individuos) y sus interacciones (lazos o conexiones) se denomina grafo, red social o simplemente \textit{red}. 
Estas estructuras son comunes en muchas áreas del conocimiento, entre ellas: finanzas (e.g., relaciones de alianzas entre países), epidemiología (e.g., relaciones de contagio de una enfermedad infecciosa), ciencias sociales (e.g., relaciones de colaboración legislativa), entre otras. 
En particular, en ciencias políticas son numerosas las investigaciones que enfatizan en la pertenencia de la perspectiva estructural como vía para comprender la influencia e interdependencia entre acciones, actores, grupos e instituciones inmersos en un sistema político específico \cite[e.g.,][]{knoke1990political,ward2011network,zech2016social,alschner2023network,de2023networked}.

Los métodos estadísticos para analizar redes permiten explicar cómo las relaciones están asociadas con la estructura y el comportamiento del sistema en el que los actores se encuentran inmersos \cite[ver][para más detalles sobre el análisis estadístico de redes sociales]{kolaczyk}. 
Desde esta perspectiva, investigaciones recientes destacan los métodos estadísticos de redes como técnicas novedosas para proveer evidencia empírica sobre la formación estratégica de vínculos y el comportamiento de agentes políticos conectados entre sí \cite[e.g.,][]{ward2011network,cardenas2015latin,de2023networked,zhang2023deep}. Adicionalmente, estudios contemporáneos afirman que el análisis de redes ofrece un rigor empírico superior al análisis individualista tradicional que considera la acción política como inherente a los actores y no al vínculo entre ellos \cite[e.g.,][]{victor2016introduction}.

En esta vía, los métodos estadísticos favorecen tres aspectos. 
Primero, resumir patrones que caracterizan la estructura de una red junto con sus entidades individuales. 
Segundo, proponer modelos probabilísticos para explicar el proceso generativo de los datos. 
Y tercero, predecir relaciones faltantes o futuras teniendo en cuenta las propiedades estructurales de la red. 
Nosotros nos enfocamos en las dos primeras vías de análisis.
En particular, proponemos implementar métodos de redes para caracterizar el discurso presidencial e identificar comunidades políticas en Colombia. 
Así, planteamos una aproximación empírica a la estructura discursiva de los jefes de Estado y a la configuración de relaciones de alianza y trabajo entre figuras prominentes de la política colombiana.

Desde la perspectiva del análisis del discurso político encontramos que es un campo de investigación activo. 
En particular, en América Latina éste se ha convertido en una vía para interpretar desde la acción comunicativa el colapso de las democracias y el resurgimiento del populismo tanto de izquierda como de derecha liderado por gobernantes autoritarios \cite[ver][]{bolivar2017political,garcia2021network,oner2023populist}. 
Luego, el discurso en cualquiera de sus manifestaciones (oral o escrita) es una fuente de información que permite a los científicos sociales y humanistas reconocer intenciones persuasivas, identidades culturales y sociales, preferencias políticas, mecanismos de control de poder, entre otros rasgos latentes (características no observables) que subyacen a las actuaciones de los actores políticos \cite[e.g.,][]{ayeomoni2012pragmatic,rudkowsky2018more,espinel2019polarizacion,garifullina2021inaugural,oladayo2023language}. 
Por lo tanto, el análisis del discurso es relevante para comprender actores, sucesos, procesos y sistemas políticos en un contexto particular.

Actualmente, hallamos un acopio importante de literatura que presenta análisis cualitativos de textos políticos en diferentes escenarios. 
Desde esta perspectiva, encontramos que las intervenciones comunicativas de los jefes de Estado de Colombia ya han sido objeto de estudio en el pasado \cite[e.g.,][]{cardenas2012aparato,villarraga2012analisis,espinel2019polarizacion,rios2022discurso}. No obstante, identificamos en el contexto colombiano pocas investigaciones que involucran técnicas cuantitativas en el análisis del discurso político \cite[e.g.,][]{villarraga2012analisis}.
Hasta lo mejor de nuestro conocimiento, para Colombia no hay estudios comparativos que contrasten desde un punto de vista cuantitativo las alocuciones de posesión presidencial de diferentes mandatarios. 
En consecuencia, aprovechamos esta novedad para ilustrar el uso de métodos de redes sociales en el análisis del discurso \cite[ver][para más detalles del análisis de redes discursivas en ciencias políticas]{leifeld2017discourse,leifeld2020policy,eder2023discourse}.

Por otro lado, adoptamos el enfoque estructural para detectar comunidades políticas a través del estudio de relaciones de alianza y de trabajo entre figuras prominentes de la política colombiana. 
El estudio de este tipo de relaciones es frecuente para examinar configuraciones de élites políticas en diferentes ámbitos \cite[e.g.,][]{yun2014climate,haim2016alliance,bai2023web,blair2023strategic}. 
Aunque ambas relaciones ya han sido objeto de pesquisa en escenarios internacionales, en el caso de la política colombiana los estudios al respecto aún son escasos.
Consideramos que investigaciones en esta dirección favorecen el análisis cuantitativo de las dinámicas de la política colombiana y ofrecen herramientas metodológicas para los estudiosos de este campo. 
Por ello, en la misma línea de \cite{garcia2021network}, utilizamos perfiles públicos para examinar las relaciones. 
Sin embargo, a diferencia de estos autores proponemos implementar un modelo de bloques estocásticos para la detección de comunidades políticas \cite[ver][]{nowicki2001estimation,latouche2011overlapping,abbe2017community}, lo que permite cuantificar directamente la incertidumbre asociada con el agrupamiento, además de llevar a cabo las tareas típicas de un modelo probabilístico como predicción/imputación.

El documento se estructura como sigue. 
En la Sección \ref{Sec2} presentamos el análisis de los discursos de posesión presidencial. 
En particular, hacemos la descripción del vocabulario, un análisis de sentimientos y cuantificamos la relación entre palabras a través de estadísticas de redes. 
En la Sección \ref{Sec3} exponemos las relaciones de trabajo y alianzas de actores políticos. 
Específicamente, definimos y describimos ambas redes e implementamos el modelo de bloques estocásticos para detectar comunidades políticas. 
Finalmente, en la Sección \ref{Sec4} planteamos conclusiones y algunas oportunidades para trabajos futuros.

%%%
\section{Discursos de posesión presidencial} \label{Sec2}

Una forma de discurso político es la alocución de posesión presidencial. Acto inaugural del gobierno entrante que es de interés en el campo de las humanidades y ciencias sociales porque ofrece evidencia sobre las estrategias discursivas que utilizan los mandatarios para influir en el pensamiento de otros \citep{villarraga2012analisis}. 
En este trabajo comparamos los discursos de posesión de los últimos cuatro presidentes de Colombia \textit{Álvaro Uribe Vélez} (2002--2010), \textit{Juan Manuel Santos Calderón} (2010--2018), \textit{Iván Duque Márquez} (2018--2022) y \textit{Gustavo Francisco Petro Urrego} (2022--2026). Unificamos en un sólo discurso las dos alocuciones de Uribe. También lo hacemos para Santos. Decidimos unificar los discursos de estos presidentes debido 
%a la similitud de sus políticas de gobierno en sus dos cuatrienios y al ejercicio de su mandato en periodos consecutivos. 
a que no encontramos diferencias sustanciales al realizar el análisis por separado.
Los discursos provienen de la sala de prensa de la Presidencia de la República de Colombia \footnote{\url{https://presidencia.gov.co/prensa/noticias}.} en idioma español.

En el mismo espíritu de \cite{blei2003latent}, tomamos cada palabra (\textit{token}, \textit{término} o \textit{unigrama}) del discurso (\textit{documento}) como unidad básica de análisis. Asumimos el enfoque de bolsa de palabras como vía para incorporar elementos de redes en el análisis del discurso \citep[ver][]{qader2019overview}. En este sentido, preprocesamos el texto para representarlo sólo con los tokens que reflejan el significado central de cada documento. El preprocesamiento consiste en eliminar caracteres especiales, suprimir números y puntuación, transformar a minúsculas, remover texto \textit{unicode}, y excluir lenguaje sin relevancia semántica (\textit{stop-words}, i.e., artículos, preposiciones, pronombres, adjetivos comunes, entre otros).

No utilizamos lematización \citep[e.g.,][]{porter2006algorithm} porque los morfemas de las palabras (en español) nos permiten evidenciar diferencias discursivas entre los mandatarios. Por ejemplo, observamos que en los discursos de Santos y Duque sólo se utiliza \textit{colombianos} como término genérico para dirigirse al pueblo (ver Figura \ref{fig:wordcloud}). No obstante, en la alocución de Petro sobresale la distinción de género al incorporar en su discurso la palabra \textit{colombianas}. Luego, los morfemas de género revelan que la alocución del actual presidente difiere de sus antecesores en lo que respecta a la inclusión y reconocimiento de lo femenino en el futuro social, cultural y político de la nación. Todo el código para garantizar la reproducción de los resultados se encuentra disponible en \url{https://github.com/Klealp/Analisis-de-Redes-Sociales} y \url{https://github.com/Isagudelogal/Analisis-de-redes-sociales}.

\subsection{Descripción del vocabulario y análisis de sentimientos}

El discurso de posesión de Álvaro Uribe está compuesto por 6284 palabras, el de Juan Manuel Santos por 8321, el de Iván Duque por 5154 y el de Gustavo Petro por 3840. Luego del preprocesamiento, el número de palabras por documento se reduce a 3039, 3466, 2110 y 1597, de las cuales son distintas el $62.8\%$, $52.9\%$, $61.2\%$ y $64.6\%$, respectivamente. Por lo tanto, Juan Manuel Santos es quien utiliza un vocabulario más reducido en su acción comunicativa inaugural. Al comparar los términos que se utilizan en los cuatro documentos, encontramos que sólo el $3.33\%$ son coincidentes en los discursos de los cuatro mandatarios. Algunos de los términos compartidos son \textit{pueblo}, \textit{Estado}, \textit{riqueza}, \textit{violencia}, \textit{educación} y \textit{salud}.

\begin{figure}[H]
\centering
    \subfigure[Álvaro Uribe (2002--2010)]{
    \centering
        \includegraphics[scale=0.50]{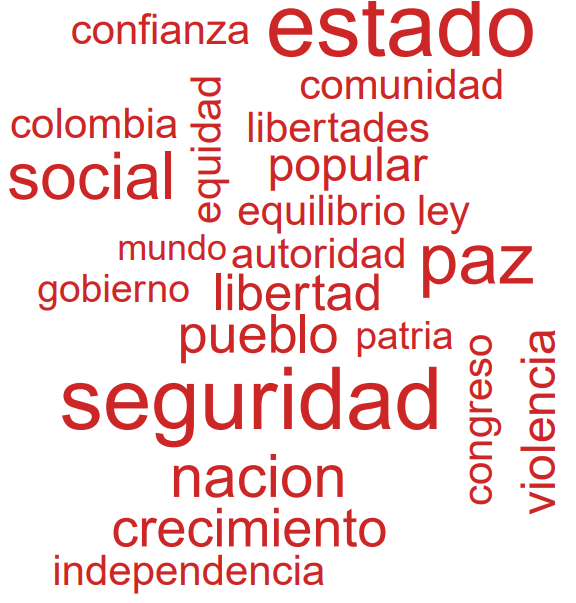}
        }
    \subfigure[Juan Manuel Santos (2010--2018)]{
    \centering
        \includegraphics[scale=0.55]{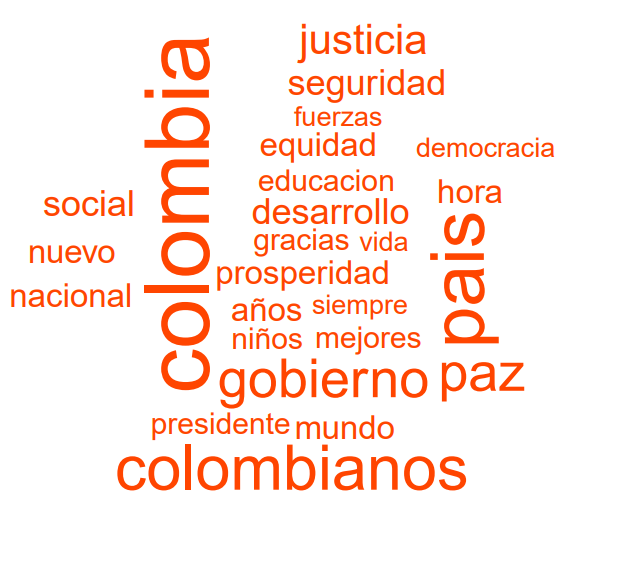}
        }
    \subfigure[Iván Duque (2018--2022)]{
    \centering
        \includegraphics[scale=0.62]{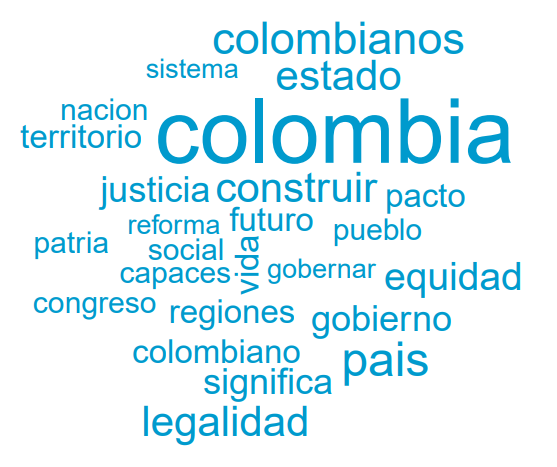}
        }
    \subfigure[Gustavo Petro (2022--2026)]{
    \centering
        \includegraphics[scale=0.65]{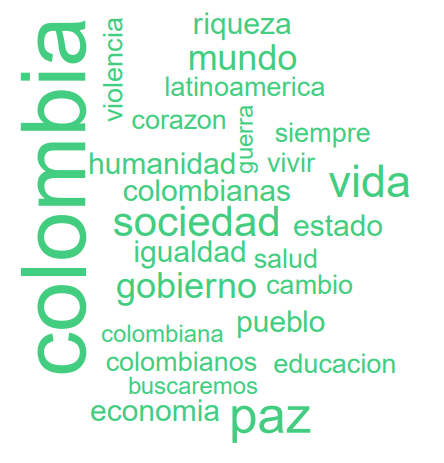}
        }
    \caption{\textit{Nube de palabras del discurso de toma de posesión presidencial. Cada ilustración muestra un máximo de 25 palabras de cada alocución. El tamaño de las palabras es proporcional a su frecuencia en el documento.}}
    \label{fig:wordcloud}
\end{figure}

Identificamos las palabras más frecuentes en cada discurso (ver Figura \ref{fig:wordcloud}). \textit{Colombia} es la palabra más popular en las alocuciones de Santos, Duque y Petro. Por otro lado, evidenciamos que la palabra \textit{paz} no es de las más usuales en la intervención de Duque. El discurso de Petro se distingue por términos como \textit{vida}, \textit{sociedad}, \textit{paz} y \textit{gobierno}. La alocución de Duque enfatiza en palabras como \textit{legalidad}, \textit{país} y \textit{construir}. Por su parte, Santos alude de forma recurrente a términos como \textit{país}, \textit{gobierno}, \textit{paz} y \textit{justicia}. Uribe utiliza de manera reiterativa términos como \textit{seguridad}, \textit{estado}, \textit{paz} y \textit{social}.

Siguiendo a \cite{haselmayer2017sentiment}, usamos diccionarios predefinidos para identificar polaridades positivas y negativas en cada documento. En la literatura existen diferentes diccionarios (léxicos de sentimientos) para el análisis textual \citep[e.g.,][]{jockers2017package,khoo2018lexicon}. La elección o creación de un diccionario para examinar sentimientos en un texto, depende del interés del investigador y del contexto de aplicación \citep[e.g.,][]{rice2021corpus}.

\begin{figure}[H]
\centering
    \subfigure[Álvaro Uribe (2002--2010)]{
    \centering
        \includegraphics[scale=0.55]{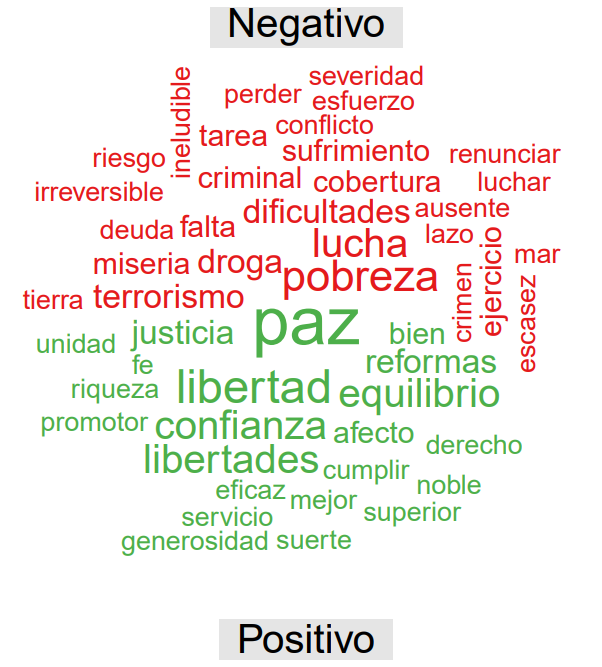}
        }
    \subfigure[Juan Manuel Santos (2010--2018)]{
    \centering
        \includegraphics[scale=0.55]{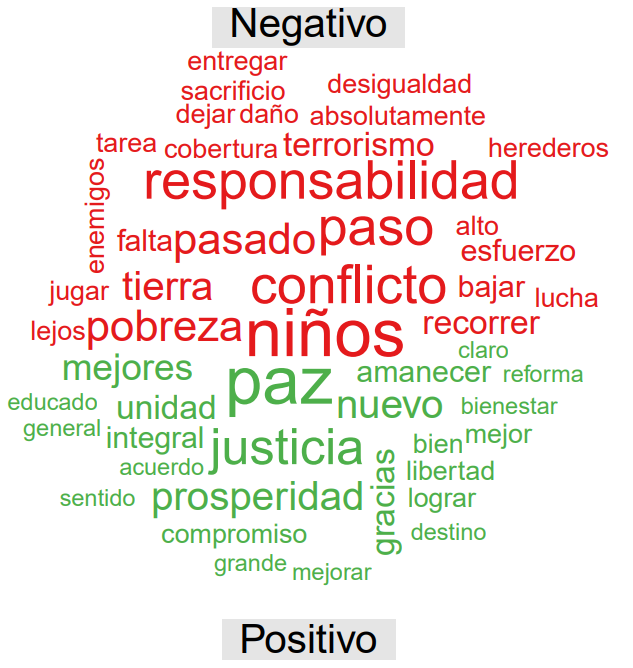}
        }
    \subfigure[Iván Duque (2018--2022)]{
    \centering
        \includegraphics[scale=0.60]{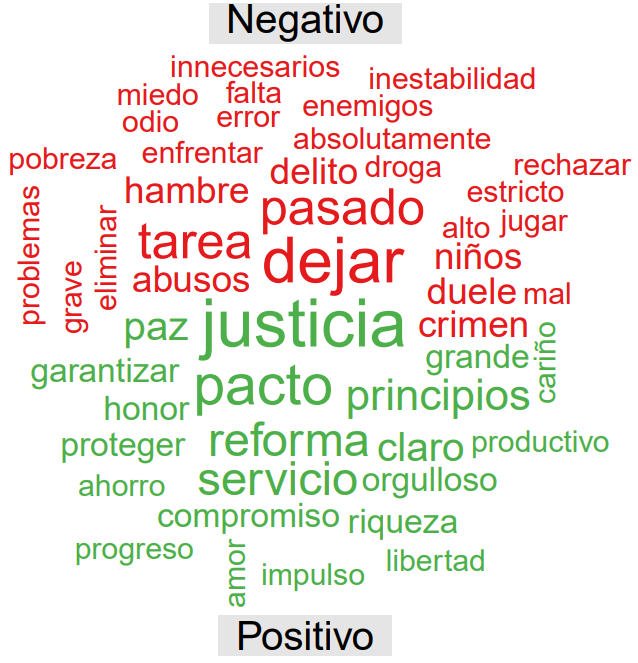}
        }
    \subfigure[Gustavo Petro (2022--2026)]{
    \centering
        \includegraphics[scale=0.63]{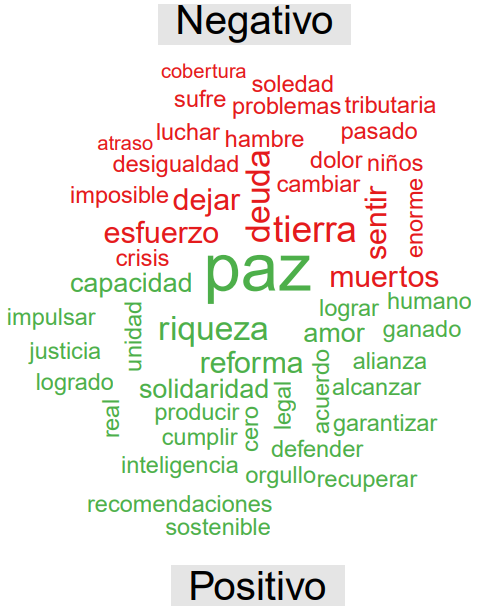}
    }
    \caption{\textit{Nube de palabras del discurso de toma de posesión presidencial. Cada ilustración muestra un máximo de 25 palabras de cada alocución. El tamaño de las palabras es proporcional a su frecuencia en el documento.}}
    \label{fig:senti}
\end{figure}

En la Figura \ref{fig:senti}, observamos que la palabra \textit{paz} es el término positivo más frecuente de los discursos de Uribe, Santos y Petro. Además, identificamos la palabra \textit{reforma} como término común a los cuatro discursos. Este último es un término de uso positivo probablemente porque en el contexto de la alocución inaugural éste es útil para demarcar una vía de cambio frente a aspectos sociales, políticos o económicos que el nuevo mandatario percibe y exhibe como perjudiciales o necesarios para el pueblo o el Estado. A través del anuncio de reformas los mandatarios manifiestan a los ciudadanos la reorganización sistemática de las condiciones y sistemas existentes \citep[][pág. 58]{sutterlutti2023reform} y enfatizan en las diferencias entre el gobierno entrante y sus antecesores o detractores.

Adicionalmente, observamos que en el discurso de los cuatro mandatarios predominan términos de carácter social (ver Figura \ref{fig:senti}). No obstante, Petro incorpora términos económicos y ambientales como \textit{sostenibilidad}. Por otro lado, en el discurso de Santos sobresalen palabras negativas como \textit{responsabilidad}, \textit{conflicto} y \textit{niños}. Estas palabras tienen una polaridad negativa quizá porque en el segundo mandato de Santos se popularizó la discusión sobre el fin del reclutamiento forzado de menores de edad y se atribuyó esta responsabilidad al Estado \citep[e.g.,][]{Martuscelli2018}. A través de los términos negativos, observamos que los cuatro mandatarios manifiestan una imperiosa necesidad de combatir la violencia, el conflicto armado, la pobreza y desigualdad en el país. Adicionalmente, reconocemos que en las palabras negativas de los discursos de Uribe y Duque está presente el término \textit{droga}. Este término refleja la puesta de estos dos mandatarios por la lucha contra el narcotráfico \citep{bustamante2020practicas, guerra2021desecuritizacion}.

En el discurso de Uribe sobresalen sentimientos de \textit{ausencia}, \textit{libertad}, \textit{confianza}, \textit{justicia} y \textit{generosidad} (ver Figura \ref{fig:senti}). Santos también exalta la \textit{justicia} y \textit{libertad}. Sin embargo, este último también hace énfasis en \textit{responsabilidad}, \textit{bienestar}, \textit{prosperidad} y \textit{compromiso}. Duque alude a sentimientos como el \textit{odio}, \textit{miedo}, \textit{rechazo}, \textit{dolor}, \textit{abuso}, \textit{cariño}, \textit{amor}, \textit{servicio}, \textit{orgullo}, \textit{justicia} y \textit{libertad}. Por su parte Petro, alude a sentimientos como \textit{soledad}, \textit{dolor}, \textit{esfuerzo}, \textit{sufrimiento}, \textit{amor}, \textit{orgullo}, \textit{justicia}, \textit{unidad} y \textit{solidaridad}. En este sentido, todos los mandatarios utilizan sentimientos antagónicos para trasmitir su mensaje al pueblo colombiano. No obstante, observamos un vocabulario más amplio en términos de emociones en el discurso de Duque y Petro. Adicionalmente, identificamos en los discursos de Uribe y Santos un mayor número de palabras que aluden a sentimientos negativos (ver Tabla \ref{tab:sentimientos}).

\begin{table}[H]
\begin{center}
\begin{tabular}{lcccc}
\hline
Polaridad & Uribe & Santos & Duque & Petro \\ \hline
Positivo    & 32.7  & 22.2   & 37.8  & 36.7 \\
Negativo    & 67.3  & 77.8   & 62.2  & 63.3 \\ \hline
\end{tabular}
\caption{Polaridad de cada discurso. Cifras expresadas en puntos porcentuales.}
\label{tab:sentimientos}
\end{center}
\end{table}

\subsection{Relación entre palabras: bigramas y skip-gramas}

\begin{figure}[H]
\centering
    \subfigure[Álvaro Uribe (2002--2010)]{
    \centering
        \includegraphics[scale=0.59]{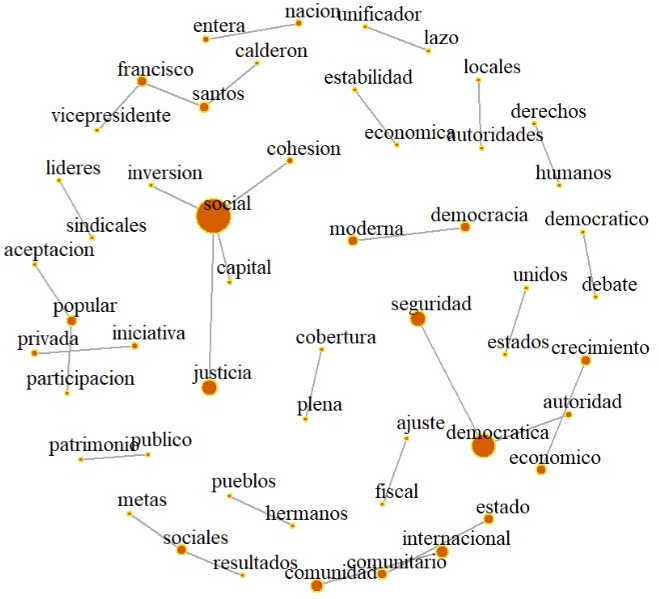}
        }
    \subfigure[Juan Manuel Santos (2010--2018)]{
    \centering
        \includegraphics[scale=0.59]{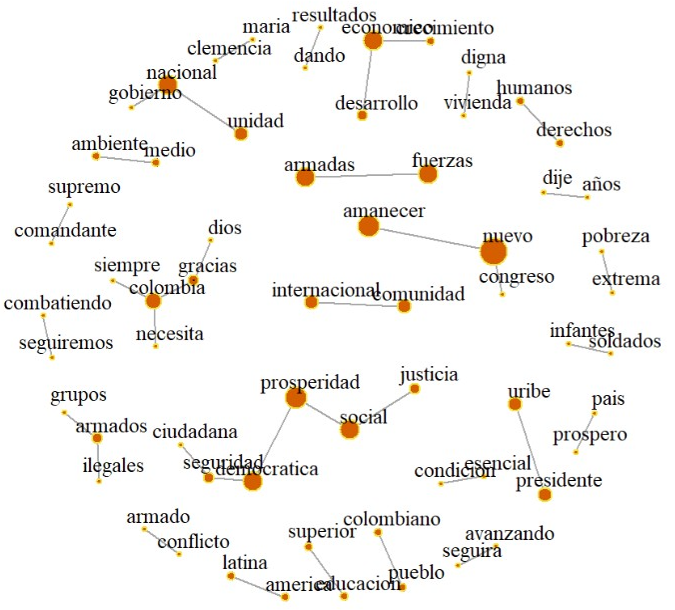}
        }
    \caption{\textit{Grafo no dirigido construido a partir de bigramas con un peso mayor o igual a dos en las aristas. Representa las redes de palabras del discurso de toma de posesión presidencial de Álvaro Uribe y Juan Manuel Santos.}}
    \label{fig:re_big}
\end{figure}

\begin{figure}[!htb]
\centering
    \subfigure[Álvaro Uribe (2002--2010)]{
    \centering
        \includegraphics[scale=0.47]{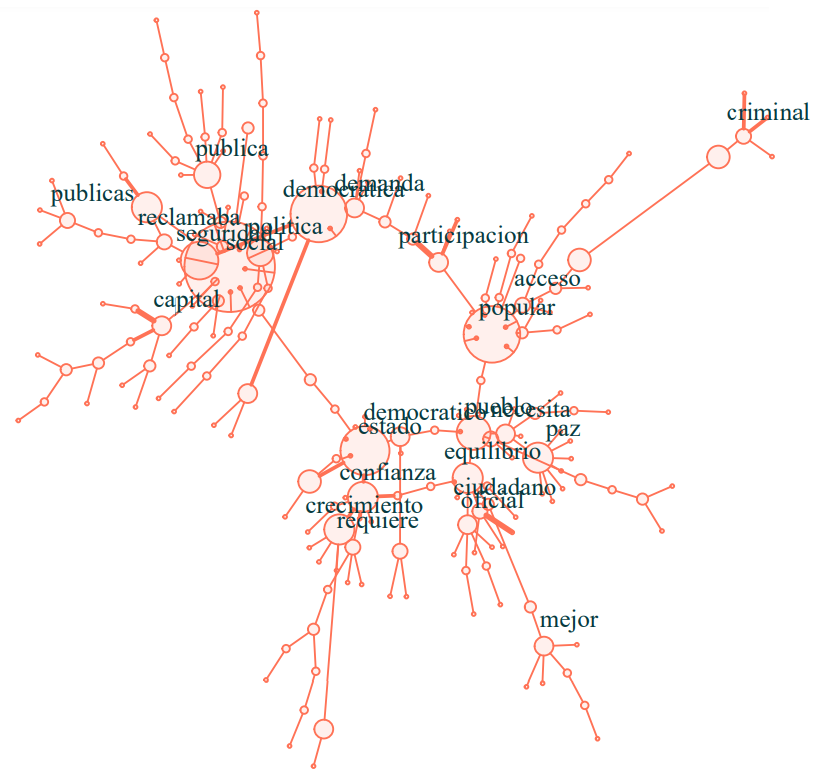}
        }
    \subfigure[Juan Manuel Santos (2010--2018)]{
    \centering
        \includegraphics[scale=0.47]{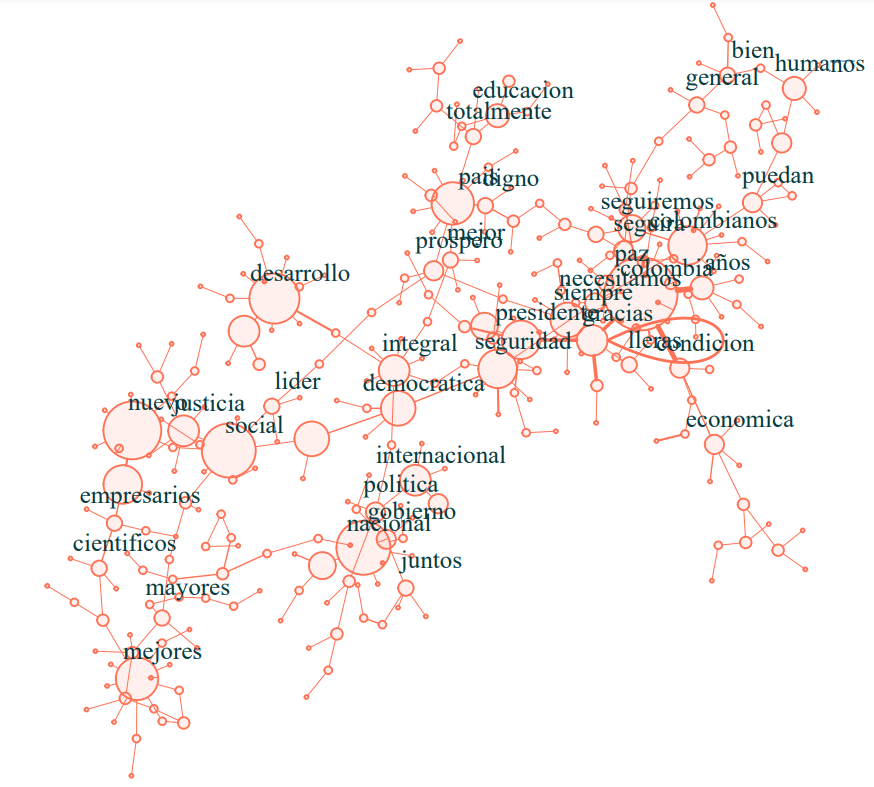}
        }
    \subfigure[Iván Duque (2018--2022)]{
    \centering
        \includegraphics[scale=0.49]{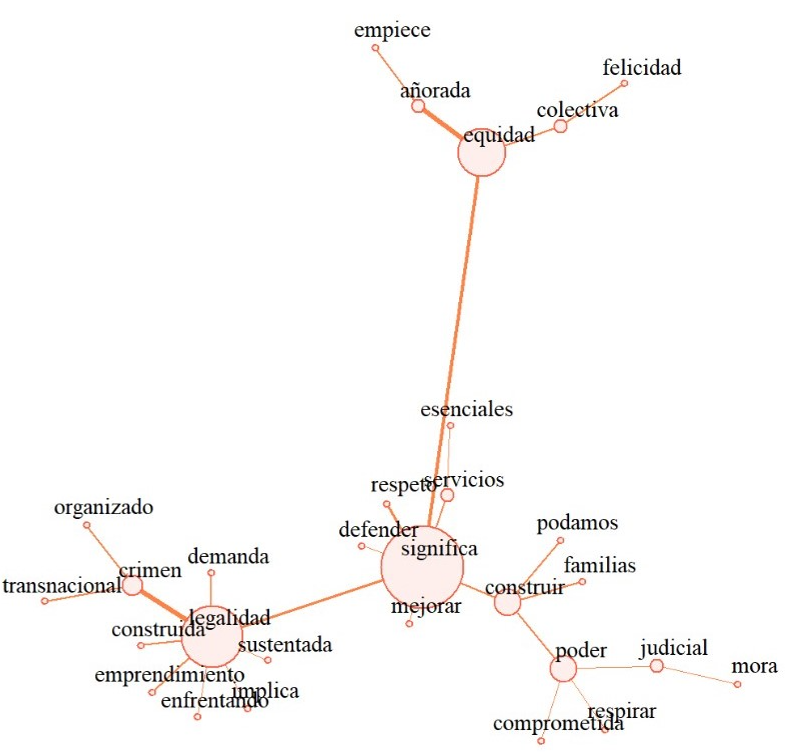}
        }
    \subfigure[Gustavo Petro (2022--2026)]{
    \centering
        \includegraphics[scale=0.49]{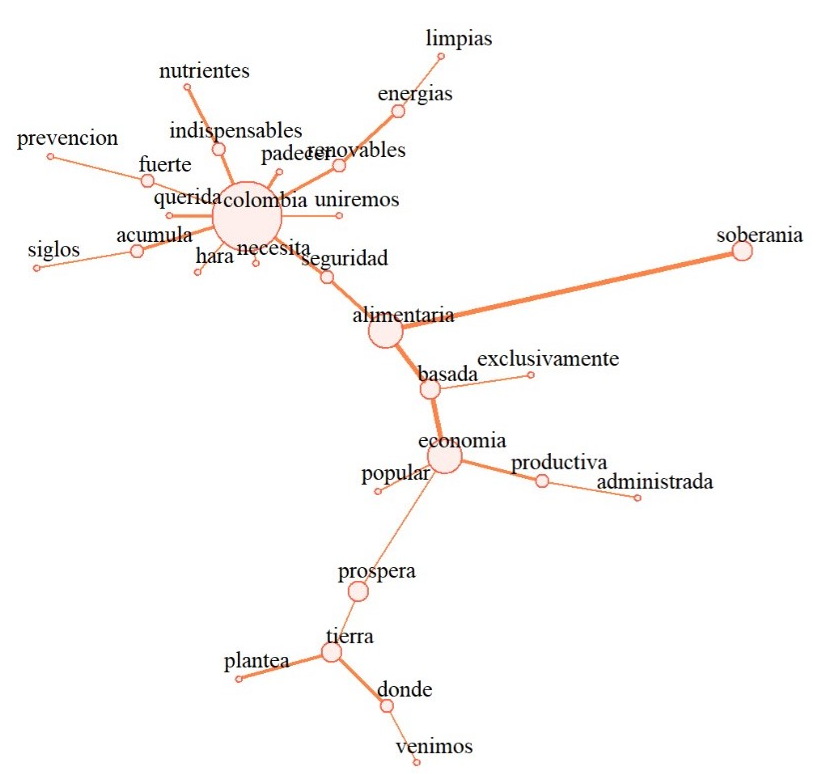}
        }
    \caption{\textit{Grafo no dirigido construido a partir de los bigramas. Representa la componente gigante (subgrafo más grande de cada red) de las redes de palabras de los discursos de toma de posesión presidencial de los cuatro mandatarios.}}
    \label{fig:com_big}
\end{figure}

Para cada discurso construimos un grafo no dirigido donde los vértices representan palabras y las aristas indican relaciones entre palabras (ver Figura \ref{fig:re_big} y \ref{fig:com_big}). Dos palabras están conectadas si y solo si son consecutivas en el discurso (\textit{bigrama}) o si son consecutivas pero están separadas por una palabra intermedia (\textit{skip-grama}). El grafo está ponderado por la frecuencia observada, i.e., el grosor de sus aristas es proporcional al número de veces que ocurre la pareja de términos en el documento. Similarmente con el tamaño de los nodos.

\newpage

A partir de las frecuencias observadas de los bigramas es posible identificar las temáticas centrales en cada discurso. Por ejemplo, en sus respectivas alocuciones, Uribe y Santos enfatizan temas políticos y económicos propios de su plan de gobierno. Como se observa en la Figura \ref{fig:re_big}, algunos de los bigramas que caracterizan el discurso de Uribe son \textit{justicia social}, \textit{seguridad democrática}, \textit{estado comunitario} e \textit{iniciativa privada}, y en el caso de Santos son \textit{unidad nacional}, \textit{desarrollo económico}, \textit{prosperidad democrática} y \textit{nuevo amanecer}. Las expresiones \textit{justicia social}, \textit{comunidad internacional} y \textit{crecimiento económico} son comunes a los dos mandatarios. 
Así mismo, como se observa en la Figura \ref{fig:com_big},
en el caso del discurso de Iván Duque sobresalen combinaciones de palabras que enfatizan aspectos sociales y de justicia. Los bigramas más populares de Duque son \textit{equidad añorada}, \textit{legalidad crimen} y \textit{poder judicial}. Finalmente, en el caso de Petro predominan bigramas que enfatizan componentes ambientales y económicos. En su alocución son comunes las expresiones \textit{energías renovables}, \textit{energías limpias}, \textit{soberanía alimentaria} y \textit{economía productiva}.

En relación con los nodos más populares, observamos que Petro centra su discurso en la palabra \textit{Colombia} derivando de ella una descripción del país en lo que respecta a sus necesidades ambientales y económicas (ver Figura \ref{fig:com_big}). En el discurso de Duque se revelan tres nodos importantes, a saber, \textit{equidad}, \textit{legalidad} y \textit{significa}. En el caso de Uribe algunos de los nodos importantes son \textit{social}, \textit{seguridad}, \textit{democrática} y \textit{justicia}. Las palabras que representan algunos de los nodos más importantes en el discurso de Santos son \textit{fuerzas}, \textit{armadas} y \textit{prosperidad} (ver Figura \ref{fig:re_big}).

\begin{table}[!htb]
\scriptsize
\centering
\begin{tabular}{lclclclc}
\hline
\multicolumn{2}{c}{\textbf{Uribe}}                 & \multicolumn{2}{c}{\textbf{Santos}}                & \multicolumn{2}{c}{\textbf{Duque}}                 & \multicolumn{2}{c}{\textbf{Petro}}                 \\ \hline
\multicolumn{1}{c}{\textbf{Palabra}} & \textbf{cp} & \multicolumn{1}{c}{\textbf{Palabra}} & \textbf{cp} & \multicolumn{1}{c}{\textbf{Palabra}} & \textbf{cp} & \multicolumn{1}{c}{\textbf{Palabra}} & \textbf{cp} \\ \hline
Seguridad                            & 1.00        & Colombia                             & 1.00        & Colombia                             & 1.00        & Colombia                             & 1.00        \\
Social                               & 0.94        & Paz                                  & 0.67        & Significa                            & 0.91        & Mujeres                              & 0.28        \\
Democrática                          & 0.80        & País                                 & 0.30        & Equidad                              & 0.89        & Empieza                              & 0.27        \\
Justicia                             & 0.65        & Hora                                 & 0.29        & Legalidad                            & 0.62        & Rincón                               & 0.27        \\
Estado                               & 0.48        & Gracias                              & 0.28        & Construir                            & 0.35        & Soñamos                              & 0.27        \\
Paz                                  & 0.44        & Dios                                 & 0.25        & Gobernar                             & 0.35        & Latinoamérica                        & 0.19        \\
Cohesión                             & 0.28        & Equidad                              & 0.23        & Necesita                             & 0.31        & Seguridad                            & 0.18        \\
Autoridad                            & 0.27        & Necesita                             & 0.21        & Futuro                               & 0.24        & Acumula                              & 0.18        \\
Reclamaba                            & 0.27        & Siempre                              & 0.20        & Pacto                                & 0.21        & Padecer                              & 0.18        \\
Demanda                              & 0.27        & Visión                               & 0.18        & Jovenes                              & 0.21        & Corazón                              & 0.17        \\ \hline
\end{tabular}
\caption{\textit{Medidas de centralidad propia de los vértices de la red (cp). Corresponden al top 10 de las palabras más importantes en cada una de las redes de palabras de los discursos de los cuatro mandatarios. Las redes incorporan bigramas y skip-gramas.}}
\label{tab:cp}
\end{table}

En la Tabla \ref{tab:cp} mostramos el top 10 de las palabras más importantes en cada alocución de acuerdo con la medida de centralidad propia (cp) de los nodos que componen las redes construidas a partir de los bigramas y skip-gramas. Esta medida permite identificar las palabras de mayor importancia teniendo en cuenta su relación con otras palabras \citep[ver][para más detalles]{ward2011network,kolaczyk}. Los resultados confirman que el nodo más importante de Petro es el que está representado por la palabra \textit{Colombia}. Este término también es el más importante en los discursos de Santos y Duque. Mientras que en el caso de Uribe el nodo más importante de la red es \textit{seguridad}.

La palabra \textit{paz} hace parte del top 10 de las palabras más importantes en el discurso de Uribe y Santos, aunque tiene mayor preponderancia (centralidad propia más alta) en la alocución de Santos. La palabra \textit{equidad} es de las más importantes en la intervención de Santos y Duque. Sin embargo, la medida de centralidad propia del nodo de equidad en la red del discurso de Duque es mayor en comparación con la medida calculada a partir de la red de la alocución de Santos. Por otro lado, \textit{mujeres} es el segundo nodo con la medida de centralidad propia más alta en la red correspondiente al discurso de Petro. Esto último confirma un discurso centrado en la inclusión de género. En el caso de Duque evidenciamos un discurso con un nodo importante en los \textit{jóvenes}. Adicionalmente, reconocemos en el discurso de Santos el vínculo del gobierno entrante con la iglesia. Uno de los nodos más importantes de la red de palabras de este mandatario es \textit{Dios}. Vale la pena citar que la iglesia católica asume el papel de ente organizador de las delegaciones de víctimas en el proceso de paz Santos-FARC, convirtiéndose en una institución clave para culminar el proceso con éxito \citep{brett2018role}.

Por otro lado, también analizamos la cohesión (conectividad entre palabras) en los discursos a través de estadísticas como la distancia geodésica media, media del grado, desviación del grado, número del clan, densidad, transitividad y asortatividad sobre las componentes gigantes de los bigramas, y bigramas y skip-gramas de cada uno de los documentos (ver Tabla \ref{tab:descrip}). Nuevamente referimos al lector al libro de \cite{kolaczyk}  para más detalles acerca de estas estadísticas descriptivas de las redes.

\begin{table}[!htb]
\scriptsize
\centering
\begin{tabular}{lrrrrrrrr}
\cline{2-9}
                                   & \multicolumn{4}{c}{\textbf{Bigramas}}                                                                                                              & \multicolumn{4}{c}{\textbf{Bigramas/Skip-gramas}}                                                                                                           \\ \cline{2-9} 
                                   & \multicolumn{1}{c}{\textbf{Uribe}} & \multicolumn{1}{c}{\textbf{Santos}} & \multicolumn{1}{c}{\textbf{Duque}} & \multicolumn{1}{c}{\textbf{Petro}} & \multicolumn{1}{c}{\textbf{Uribe}} & \multicolumn{1}{c}{\textbf{Santos}} & \multicolumn{1}{c}{\textbf{Duque}} & \multicolumn{1}{c}{\textbf{Petro}} \\ \hline
\textbf{Distancia geodésica media} & 9.89                               & 12.30                               & 5.16                               & 4.96                               & 9.89                               & 7.13                                & 5.16                               & 8.61                               \\
\textbf{Media del grado}           & 2.04                               & 2.08                                & 1.93                               & 1.93                               & 2.60                               & 2.72                                & 2.38                               & 2.31                               \\
\textbf{Desviación del grado}      & 1.92                               & 1.87                                & 1.79                               & 1.75                               & 2.48                               & 3.13                                & 2.17                               & 2.25                               \\
\textbf{Número de clan}            & 3.00                               & 3.00                                & 2.00                               & 2.00                               & 3.00                               & 4.00                                & 3.00                               & 3.00                               \\
\textbf{Densidad}                  & 0.01                               & 0.01                                & 0.07                               & 0.07                               & 0.00                               & 0.00                                & 0.00                               & 0.00                               \\
\textbf{Transitividad}             & 0.00                               & 0.01                                & 0.00                               & 0.00                               & 0.08                               & 0.06                                & 0.06                               & 0.06                               \\
\textbf{Asortatividad}             & -0.23                              & -0.23                               & -0.37                              & -0.36                              & 0.01                               & -0.05                               & -0.04                              & -0.08                              \\ \hline
\end{tabular}
\caption{\textit{Estadísticas descriptivas de las componentes gigantes de los bigramas, y bigramas y skip-gramas de cada uno de los discursos.}}
\label{tab:descrip}
\end{table}

En la Tabla \ref{tab:descrip} evidenciamos existencia de patrones transitivos en los bigramas de la red de palabras correspondiente al discurso de Santos. No obstante, la frecuencia relativa con la que las triplas de palabras forman ciclos es baja. Por ejemplo, en la Figura \ref{fig:re_big} observamos que la tripla prosperidad-social-justicia no forma un ciclo transitivo puesto que no hay conexión entre prosperidad y justicia. La ausencia de patrones transitivos en los discursos es indicador de una falta de conexión entre parejas de bigramas. Por lo tanto, notamos que la transitividad es una medida útil para determinar si es o no adecuado examinar relaciones entre términos (de acuerdo con el contexto) o hacer procesos deductivos cuando el texto se representa a partir de pares de términos. Además, consideramos que el uso de esta estadística puede contribuir a ampliar la discusión acerca de las limitaciones que tiene el enfoque de bolsa de palabras en el procesamiento del lenguaje natural. En este sentido, destacamos la transitividad como una medida útil para examinar la pérdida de estructura gramatical cuando se seleccionan solo algunos términos para representar el documento original \cite[ver][]{rudkowsky2018more}.

Al incorporar bigramas y skip-gramas en el cálculo de las estadísticas mencionadas, observamos que la evidencia de patrones transitivos aumenta para las redes de palabras de los discursos de los cuatro mandatarios (ver Tabla \ref{tab:descrip}). Es decir, evidenciamos una mayor conexión entre triplas de palabras. Sin embargo, no estamos seguros que haya una mayor posibilidad de encontrar información no-explícita en los discursos, así como tampoco que estas nuevas relaciones mantengan el significado semántico del texto original. Vale la pena aclarar que los procesos deductivos semánticos que se realicen a partir de bigramas y skip-gramas no sólo deben apoyarse en las estadísticas, también deben tener sustento en el contexto político del documento objeto de análisis y en el enfoque que tenga el científico social o humanista interesado.

Respecto a la densidad, observamos que para el conjunto de bigramas y el de bigramas y skip-gramas de las redes de palabras asociadas a los discursos de los cuatro mandatarios, la frecuencia relativa de las conexiones observadas entre palabras respecto al posible número de conexiones que se podrían originar es baja o incluso nula (ver Tabla \ref{tab:descrip}). Por ejemplo, en el caso de la red de bigramas del discurso de Uribe, de todas las posibles conexiones entre palabras solo se observa el 1\% de esas conexiones. Este porcentaje aumenta para los bigramas de Duque y Petro. Consideramos que esta medida es útil en el contexto del análisis del discurso para distinguir entre aquellos documentos que puntualizan en temáticas específicas de aquellos que involucran una amplia variedad de temas. En otras palabras, suponemos que la medida de densidad se incrementa cuando los bigramas provienen de textos que se desarrollan alrededor de tópicos puntuales y se disminuye cuando los bigramas se extraen de documentos que abarcan múltiples temas sin enfatizar de manera particular en alguno de ellos. En este sentido, argumentamos que los discursos de los cuatro mandatarios comprenden un número amplio de tópicos sin una acentuación puntual en ninguno de ellos. Por otro lado, observamos que los skip-gramas no proporcionan información adicional en relación con la conexión entre palabras.

Apoyamos la idea de baja conectividad entre pares de palabras a través de la estadística media del grado. Para todos los escenarios esta estadística muestra que el número  medio de aristas incidentes en cada nodo es alrededor de dos, i.e., en promedio cada palabra seleccionada para representar el discurso se relaciona con otras dos. Además, observamos que la desviación del grado se incrementa cuando se consideran bigramas y skip-gramas. De esto último, concluimos que incorporar los skip-gramas implica tener una red más grande, pero más dispersa en términos de conectividad entre parejas de palabras.

Por otro lado, encontramos para el caso de los bigramas que las medidas de densidad, asortatividad, número del clan, y media del grado son similares para las redes de palabras de Petro y Duque, y para las redes de Santos y Uribe (ver Tabla 3). Entre pares de discursos estas medidas difieren. Por ejemplo, la medida de asortatividad de las redes de Uribe y Santos es más alta frente a la calculada para las redes de Duque y Petro. La asortatividad es negativa en la mayoría de los escenarios, excepto para la red conformada por los bigramas y skip-gramas extraídos del discurso de Uribe. 
Finalmente, notamos que la única estadística con una diferencia amplia entre los bigramas y los bigramas y skip-gramas extraídos de las palabras que componen los discursos  es la distancia geodésica media. La red correspondiente al discurso de Santos tiene la distancia geodésica medía más amplia en contraste con la calculada para las redes de palabras de los demás gobernantes.

\section{Relaciones de trabajo y alianza de actores políticos} \label{Sec3}

En esta sección analizamos los perfiles disponibles en la sección \textit{Quién es Quién} del medio de comunicación independiente \textit{La Silla Vacía}. 
A partir de estos examinamos relaciones de alianza y de trabajo entre figuras prominentes de la política colombiana.
Los datos son recopilados mediante técnicas de raspado de páginas web (\textit{web scraping}).
Sugerimos al lector interesado ver el trabajo de \citet{garcia2021network}, dado que es pionero en la conformación de redes usando estas técnicas.

Obtenemos los datos de 664 actores políticos en formato de texto plano. Dentro de los perfiles recuperados están los correspondientes a los presidentes de Colombia \textit{César Augusto Gaviria Trujillo} (1990--1994), \textit{Andrés Pastrana Arango} (1998--2002), \textit{Álvaro Uribe Vélez} (2002--2010), \textit{Juan Manuel Santos} (2010--2018), \textit{Iván Duque Márquez} (2018--2022) y el actual jefe de Estado \textit{Gustavo Francisco Petro Urrego} (2022-2026). También se encuentran políticos de amplia trayectoria como \textit{Enrique Peñalosa Londoño}, alcalde de Bogotá (2016--2019), \textit{Sergio Fajardo Valderrama}, alcalde de  Medellín (2004--2007) y gobernador de Antioquia (2012--2015), \textit{Alejandro Ordóñez Maldonado}, procurador (2009--2016) y embajador de Colombia ante la OEA (2018--2022), \textit{Juan Camilo Restrepo Salazar}, jefe del equipo negociador del gobierno con el ELN (2016--2018) y ministro de diferentes carteras en su trayectoria política (1991--1992, 1998--2000, 2010--2013), \textit{Germán Vargas Lleras}, vicepresidente de la República de Colombia (2014--2017), ministro de Vivienda, Ciudad y Territorio (2010--2012), y ministro del Interior y Justicia (2012--2013), entre otros.

Para analizar las relaciones entre figuras políticas, definimos la matriz de adyacencia de la red de trabajo como  $\boldsymbol{Y}= [y_{i,j}]$ donde $y_{i,j} =1$ si el funcionario $i$ nombró al funcionario $j$ en algún cargo público o viceversa. Mientras que para el caso de la matriz de adyacencia de la red de alianzas, asumimos que $y_{i,j}=1$ si los funcionarios $i$ y $j$ apoyan de manera conjunta la aprobación de algún proyecto de ley o si tienen membresía en el mismo partido o grupo político.

\subsection{Descripción de las redes políticas}

Construimos las componentes conexas de la red de trabajo y de alianzas. Estas componentes están constituidas por 485 y 335 de los 664 nodos/actores de las redes originales, respectivamente. Las componentes conexas de ambas redes comparten 189 nodos. Algunos de los actores políticos que hacen parte de ambas redes son Alejandro Ordóñez, Enrique Peñalosa, Sergio Fajardo, Gustavo Petro, Juan Manuel Santos y Álvaro Uribe.

\begin{table}[H]
\scriptsize
\centering
\begin{tabular}{lcc}
\cline{2-3}
\textbf{}                          & \textbf{Trabajo} & \multicolumn{1}{l}{\textbf{Alianzas}} \\ \hline
\textbf{Distancia geodésica media} & 3.308                   & 4.109                                        \\
\textbf{Media del grado}           & 5.687                   & 4.048                                        \\
\textbf{Desviación del grado}      & 10.538                  & 4.927                                             \\
\textbf{Densidad}                  & 0.012                   & 0.012                                        \\
\textbf{Transitividad}             & 0.099                   & 0.140                                        \\
\textbf{Asortatividad}             & -0.157                  & -0.227                                       \\ \hline
\end{tabular}
\caption{\textit{Estadísticas descriptivas de las componentes conexas de la red de trabajo y de alianzas.}}
\label{tab:descrip_TA}
\end{table}

En la Tabla \ref{tab:descrip_TA} presentamos las estadísticas estructurales de las componentes conexas de la red de trabajo y de alianzas. Observamos la misma densidad para ambas redes. Esta medida indica que existe una baja probabilidad (aproximadamente del 1\%) de que dos políticos cualesquiera de cada red estén relacionados de manera directa. La baja conectividad es consecuencia de la diversidad de cargos públicos, membresías y trayectoria política de los actores considerados en el análisis. Adicionalmente, la estadística de transitividad revela que son poco frecuentes las ternas de funcionarios que trabajan o realizan alianzas entre sí. Además, a través de la distancia geodésica promedio identificamos que las figuras políticas en ambas componentes están separadas unas de otras en promedio por tres o cuatro nodos aproximadamente, siendo esta cantidad menor para la red de trabajo en comparación con la red de alianzas. Por otro lado, la media del grado señala un número de conexiones promedio por nodo mayor para la red de trabajo en contraste con la red de alianzas, aunque con una mayor desviación (ver Tabla \ref{tab:descrip_TA}).

\begin{figure}[H]
\centering
    \subfigure[Red de trabajo]{
    \centering
        \includegraphics[scale=0.65]{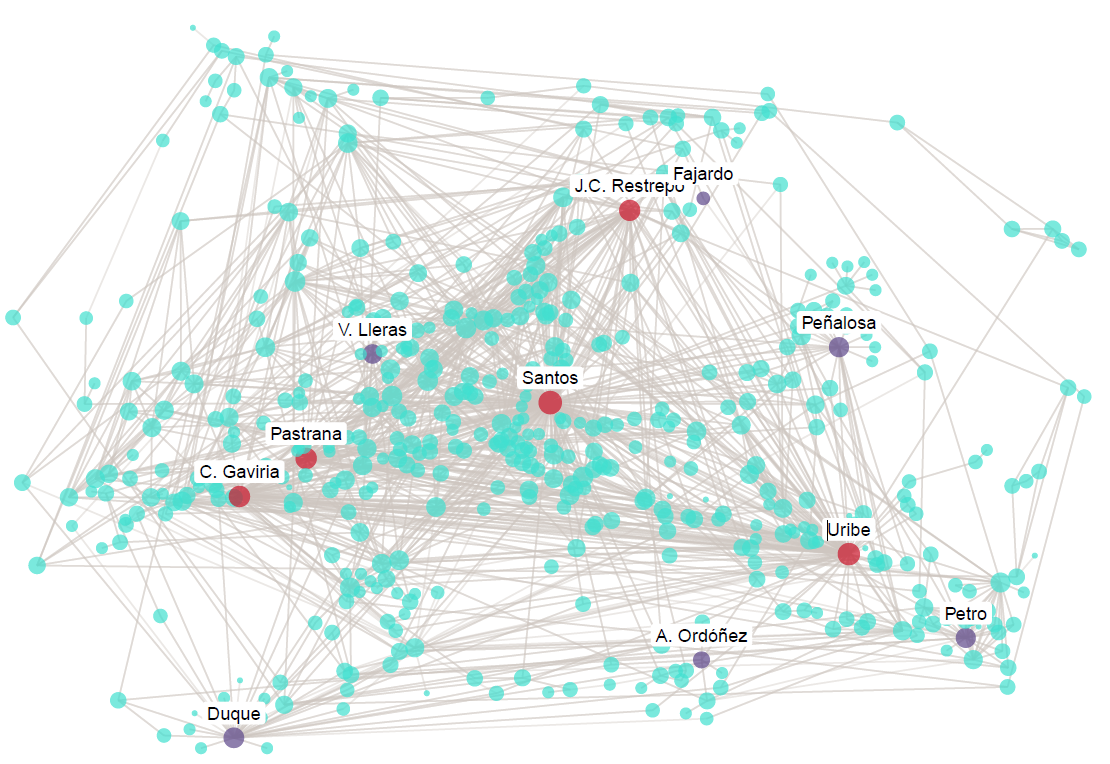}
        }
    \subfigure[Red de alianzas]{
    \centering
        \includegraphics[scale=0.65]{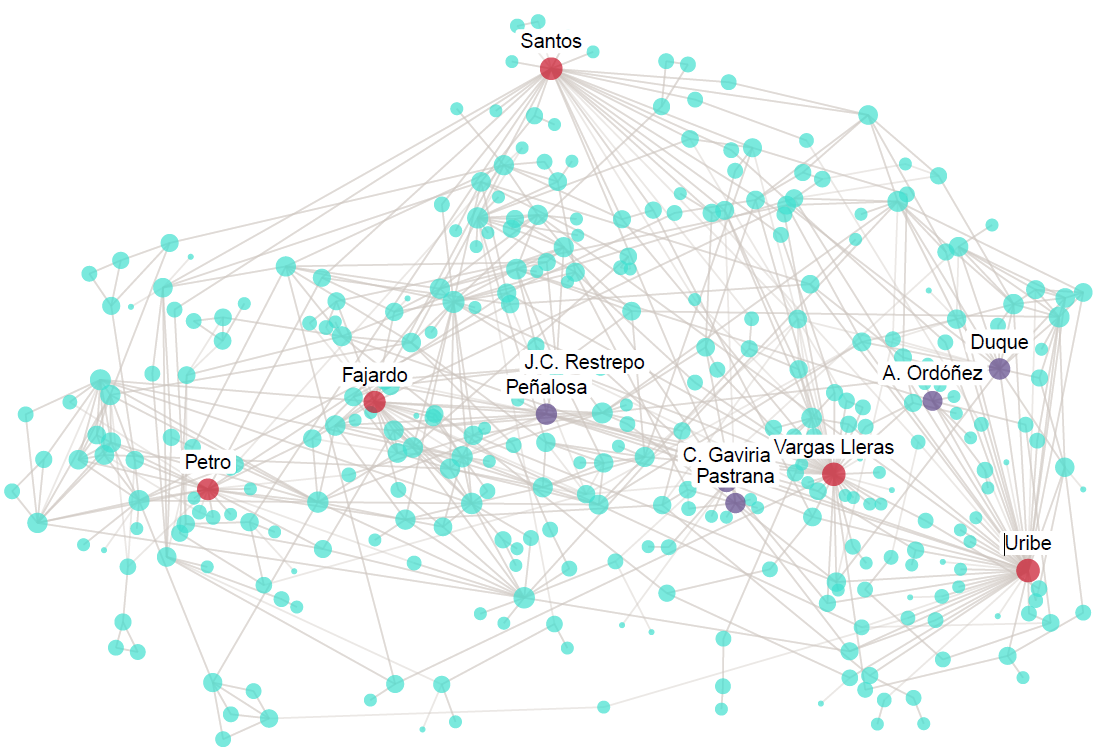}
        }
    \caption{\textit{Red de trabajo y de alianzas. En cada red el tamaño de los nodos es proporcional al grado. Los nodos en color rojo son los más importantes según la medida del grado. Los nodos en color morado corresponden a otras personalidades importantes de la política nacional.}}
    \label{fig:red_AT}
\end{figure}

En la red de trabajo, los nodos con más conexiones corresponden a funcionarios que se han desempeñado como presidentes o han ocupado algún ministerio (ver Figura \ref{fig:red_AT}). La potestad del presidente para nombrar ministros y directores de departamentos administrativos (ver artículo 189 de la constitución política de Colombia, 1991), y su relación con las diferentes ramas del poder público, hace que estos actores tengan un mayor número de relaciones de trabajo frente a los demás miembros de esta red.

En la red de trabajo evidenciamos que los nodos correspondientes a César Gaviria y Andrés Pastrana son cercanos (ver Figura \ref{fig:red_AT}). Ambos políticos revelan conexiones de trabajo similares a pesar de ser miembros de partidos antagónicos y hegemónicos de tradición (liberal y conservador, respectivamente). Tal cercanía quizá es atribuible a su participación en las diferentes iniciativas políticas de paz en Colombia \cite[ver][]{nasi2010saboteadores}. Por otro lado, observamos que Juan Manuel Santos se ubica en el centro de la red de trabajo y despliega conexiones en todas las direcciones. El expresidente Santos es una de las figuras políticas más relevantes del país, sus conexiones laborales inician públicamente en el gobierno de César Gaviria como ministro de Comercio Exterior (1991--1994). Luego, Santos ejerce como ministro de Hacienda y Crédito Público (2000--2002) en el gobierno de Andrés Pastrana, y como ministro de Defensa (2006--2009) en el gobierno de Álvaro Uribe. Posteriormente, Santos nombra a Juan Camilo Restrepo como ministro de Agricultura y Desarrollo Rural (2010--2013) en su primer mandato presidencial. Además, Restrepo también fue ministro de Minas y Energía (1991--1992) de Gaviria y ministro de Hacienda y Crédito Público (1998--2000) de Pastrana. Por lo anterior, advertimos que las relaciones laborales de Santos y Restrepo con otros miembros importantes de la política colombiana se han tejido a través del nombramiento de cargos públicos, principalmente ministeriales.

En la red de trabajo, también apreciamos que Álvaro Uribe y César Gaviria se ubican en lados opuestos de la red. No obstante, las conexiones de ambos demarcan una línea que los une (ver Figura \ref{fig:red_AT}a). Consideramos que tal conexión laboral tiene raíz en la membresía de partido. Uribe militó en el partido liberal (1977--2001) y fue senador de la República (1991--1994) por esta colectividad durante la presidencia de Gaviria. Es de resaltar que las relaciones laborales de Uribe se caracterizan por la lealtad de partido \citep[e.g.,][]{fierro2014alvaro}.

\begin{table}[!bh]
\scriptsize
\centering
\begin{tabular}{lcccc}
\cline{2-5}
\textbf{}                   & \multicolumn{2}{c}{\textbf{Trabajo}}                                     & \multicolumn{2}{c}{\textbf{Alianzas}}                                    \\ \cline{2-5} 
                            & \multicolumn{1}{l}{\textbf{Uribe}} & \multicolumn{1}{l}{\textbf{Santos}} & \multicolumn{1}{l}{\textbf{Uribe}} & \multicolumn{1}{l}{\textbf{Santos}} \\ \hline
\textbf{Grado}              & 0.18                               & 0.36                                & 0.13                               & 0.09                                \\
\textbf{Cercanía}           & 0.45                               & 0.54                                & 0.36                               & 0.32                                \\
\textbf{Intermediación}     & 0.20                               & 0.53                                & 0.30                               & 0.14                                \\
\textbf{Centralidad propia} & 0.39                               & 1.00                                & 1.00                               & 0.26                                \\ \hline
\end{tabular}
\caption{\textit{Medidas de centralidad de los vértices correspondientes a Juan Manuel Santos y a
Álvaro Uribe Vélez en la red de trabajo y de alianzas. Todas las medidas se expresan en términos relativos.}}
\label{tab:descrip_nodos}
\end{table}

En el caso de la red de alianzas, identificamos que los nodos más importantes son los correspondientes a  Álvaro Uribe, Juan Manuel Santos, Sergio Fajardo, Germán Vargas Lleras y Gustavo Petro (ver Figura \ref{fig:red_AT}b). Estos actores tienen una amplia trayectoria política. Todos han ocupado diferentes cargos públicos y han sido candidatos a la presidencia de la República en más de una ocasión. Además, sus partidos o grupos políticos de membresía (e.g., coaliciones) cuentan con representatividad en la rama legislativa del país.

En la red de alianzas, Santos se ubica en la parte superior de la red y en contraste con la red de trabajo muestra un menor número de relaciones que se despliegan desde su nodo. No obstante, todas las conexiones lo vinculan de forma indirecta con Germán Vargas Lleras, Álvaro Uribe, Gustavo Petro y Sergio Fajardo. A través de la red, no observamos alianzas directas entre estas figuras, pero vislumbramos conexión por medio de otros actores políticos.

Comparamos las medidas de centralidad (grado, cercanía, intermediación y centralidad propia) de los vértices correspondientes a Álvaro Uribe y Juan Manuel Santos en la red de trabajo y de alianzas (ver Tabla \ref{tab:descrip_nodos}). Encontramos que Uribe representa el nodo más importante de la red de alianzas mientras que Santos es el vértice más notable de la red de trabajo. Observamos que el grado, la intermediación y la centralidad propia son las métricas que presentan una diferencia más amplia entre estos dos actores en ambas redes.

Finalmente, contrastamos sus medidas de centralidad propia e identificamos que Santos tiene una mayor proporción de conexiones con nodos importantes en la red de trabajo en contraste con Uribe. No obstante, evidenciamos un patrón inverso en la red de alianzas. Esto quiere decir que, en comparación con Santos, Uribe presenta una mayor cantidad de alianzas con funcionarios que a su vez tienen convenios con muchas otras figuras políticas, mientras que las personas con las que se relaciona laboralmente no necesariamente son muy populares en la red. En otras palabras, Uribe ha destinado en puestos públicos a una cantidad considerable de personas sin mucha relevancia o trayectoria política en términos de los cargos que haya desempeñado y, sin embargo, sus alianzas se dan con los funcionarios de mayores asociaciones en términos de convenios o acuerdos, lo cual marca una clara diferencia en las estrategias y obrar político de ambos expresidentes.

\subsection{Detección de comunidades políticas}

Ajustamos un modelo de bloques estocásticos \citep{nowicki2001estimation} para las redes de trabajo y alianzas. Modelamos las conexiones sujeto/nodo de figuras prominentes del panorama político colombiano contemporáneo para explicar la forma cómo estos actores establecen relaciones entre sí. En este sentido, identificamos comunidades de actores políticos y sus respectivas probabilidades de interacción dentro y entre agrupaciones.

Utilizamos la maximización de la verosimilitud de clasificación de integración del modelo para obtener el número de particiones óptimo de cada red \citep{kolaczyk}. Así, los nodos de la red de trabajo y alianzas se agrupan en 5 y 6 particiones, respectivamente (ver Tabla \ref{tab:comuni}). Observamos para ambas redes que el grupo 1 es el que involucra el mayor número de actores políticos, i.e., el 77\% de los nodos de cada una de las redes hacen parte de la primera comunidad.

\begin{table}[H]
\scriptsize
\centering
\begin{tabular}{lccccccccccc}
\cline{2-12}
\textbf{}             & \multicolumn{6}{c}{\textbf{Trabajo}}                                        & \multicolumn{5}{c}{\textbf{Alianzas}}                          \\ \hline
\textbf{Grupo}        & \textbf{1} & \textbf{2} & \textbf{3} & \textbf{4} & \textbf{5} & \textbf{6} & \textbf{1} & \textbf{2} & \textbf{3} & \textbf{4} & \textbf{5} \\ \hline
\textbf{Tamaño}       & 377        & 17         & 2          & 70         & 9          & 10         & 258        & 16         & 3          & 46         & 12         \\ \hline
\textbf{Proporción} & 0.777      & 0.035      & 0.004      & 0.144      & 0.019      & 0.021      & 0.770      & 0.048      & 0.009      & 0.137      & 0.036      \\ \hline
\end{tabular}
\caption{\textit{Particiones propuestas para la red de trabajo y alianzas. Para cada red se muestra el número de grupos. Para cada grupo el número de nodos que los conforman (tamaño) y su respectiva proporción.}}
\label{tab:comuni}
\end{table}

\begin{figure}[H]
\centering
    \subfigure[Red de trabajo]{
    \centering
        \includegraphics[scale=0.45]{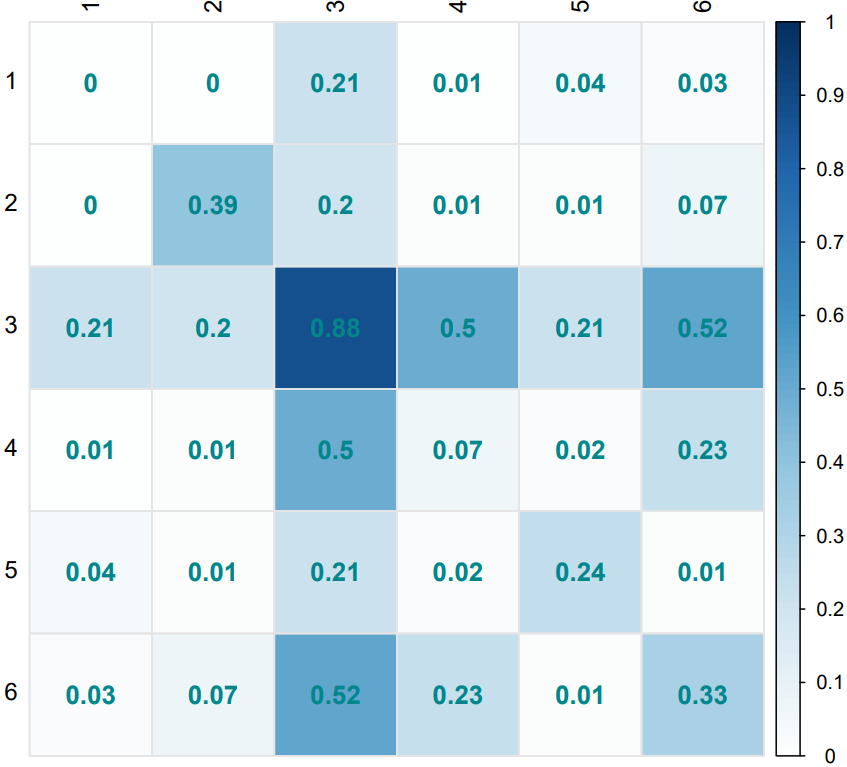}
        }
    \hfill
    \subfigure[Red de alianzas]{
    \centering
        \includegraphics[scale=0.45]{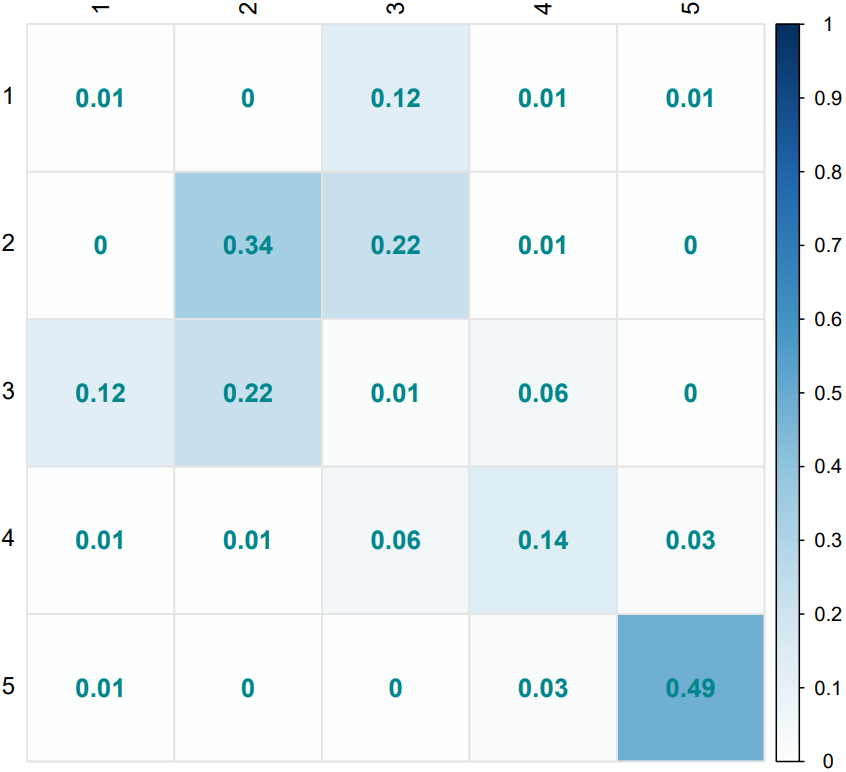}
        }
    \caption{\textit{Matrices de probabilidades de interacción entre grupos.}}
    \label{fig:interac}
\end{figure}

La estimación de las probabilidades de interacción entre comunidades revela una alta probabilidad entre los nodos que conforman el grupo 3 de la red de trabajo ($88\%$) y entre los nodos del grupo 5 de la red de alianzas ($49\%$) (ver Figura \ref{fig:interac}). El grupo 3 de la red de trabajo está conformado por Álvaro Uribe Vélez y Juan Manuel Santos (ver Figura \ref{fig:interac}). La alta probabilidad de interacción entre estos dos actores políticos es de carácter ministerial, Juan Manuel Santos fue ministro de Defensa en el segundo mandato de Uribe (2006-2010). Por lo tanto, el modelo recupera el vínculo de estos dos actores.

Por otro lado, evidenciamos una nula o baja probabilidad de interacción entre los nodos que pertenecen al grupo 1 en ambas redes. También evidenciamos una baja probabilidad de interacción entre los actores que conforman el grupo 4 de la red de trabajo ($7\%$) y entre los agentes políticos que integran el grupo 3 de la red de alianzas ($1\%$). Esto último, indica escasas relaciones entre los actores políticos que conforman estas comunidades.

Distinguimos patrones en las comunidades de la red de trabajo. Por ejemplo, identificamos que el grupo 6 está compuesto por expresidentes y funcionarios que han ejercido cargos públicos en diferentes gobiernos, tales como \textit{Juan Camilo Restrepo Salazar}.
Los miembros del grupo 5 corresponden a exalcaldes principalmente. Hacen parte de este grupo \textit{Antanas Mockus}, alcalde de Bogotá (2002--2003), y \textit{Antonio Navarro Wolf}, alcalde de Pasto (1995--1997).
El grupo 4 está conformado por ministros y vicepresidentes de los actores políticos del grupo 3. En el grupo 4 están \textit{Óscar Iván Zuluaga}, ministro de Hacienda y Crédito Público de Uribe (2007--2010), y \textit{Alejandro Gaviria}, ministro de Salud y Protección Social en el gobierno de Santos (2012--2018), entre otros. Los nodos del grupo 4 se despliegan desde el centro hacia los extremos de la misma. 
Con relación al grupo 2, identificamos que está conformado por directivos de empresas y funcionarios vinculados con el sector minero público y privado del país. Hacen parte de este grupo \textit{Amilkar Acosta Medina}, ministro de Minas y Energía (2013--2014), \textit{Beatriz Uribe Restrepo} presidenta de Mineros S.A., y \textit{Natalia Gutiérrez Jaramillo} presidenta de la Agencia Nacional de Minería (2014--Actual), entre otros.
Por último, en el grupo 1 están figuras con diferentes trayectorias políticas que han tenido algún vínculo con los expresidentes del grupo 3. Pertenecen al grupo 1, \textit{Sergio Fajardo} y \textit{Arturo Char Chaljub}, senador de la República (2006--2023), entre otros.
En la Figura \ref{fig:interac}(a) observamos que al lado izquierdo del nodo correspondiente a Santos se ubican en su mayoría directivos de empresas y expresidentes (grupos 2 y 6, respectivamente), en cambio al lado derecho se ubican exalcaldes (grupo 5). Algunos de estos últimos, con nodos cercanos al vértice de Uribe.

Por otro lado, identificamos diferencias ideológicas y de coalición en las comunidades de la red de alianzas. Por ejemplo, el grupo 5 está conformado por agentes que han hecho parte del partido Polo Democrático Alternativo (83.3\%). Tales como \textit{Jorge Enrique Robledo}, congresista (2002--2022), e \textit{Iván Cepeda} miembro actual del equipo negociador del gobierno de Colombia con el ELN. Este grupo de ubica al lado izquierdo de la red (ver Figura \ref{fig:interac}). 
El grupo 4, en su mayoría está conformado por miembros de la Alianza verde (32.6\%) y el partido Liberal Colombiano (47.8\%). Hacen parte de éste grupo \textit{Claudia López}, actual alcaldesa de Bogotá, y el expresidente \textit{Cesar Gaviria}. Los nodos del grupo 4 se localizan hacia el centro de la red. 
Las comunidades 4 y 5 están ubicadas al lado izquierdo del grupo 3. 
Este último, conformado por \textit{Álvaro Uribe}, \textit{Germán Vargas Lleras} y \textit{Juan Manuel Santos}. 
Por otro lado, el grupo 2 está conformado por funcionarios que representan la derecha en Colombia, la mayoría de ellos con membresía de partido en el Centro Democrático (93.75\%). Hacen parte de este grupo las congresistas \textit{María Fernanda Cabal} y \textit{Paloma Valencia} (2014--Actual). Los nodos del grupo 2 se localizan al lado derecho de la red, cercanos a Vargas Lleras y Uribe (miembros del grupo 3). 
Al igual que en la red de trabajo los nodos del grupo 1 están dispersos a lo largo de la red de alianzas. Algunas de las personalidades de este último grupo son \textit{Andrés Felipe Arias}, ministro de Agricultura y Desarrollo Rural (2005--2009), \textit{Armando Benedetti} congresista en varias oportunidades (2002--2022) y Embajador de Colombia en Venezuela (20022--2023), \textit{Francia Márquez} actual Vicepresidenta de la República de Colombia, entre otros.

\begin{figure}[H]
\centering
    \subfigure[Red de trabajo]{
    \centering
        \includegraphics[scale=0.64]{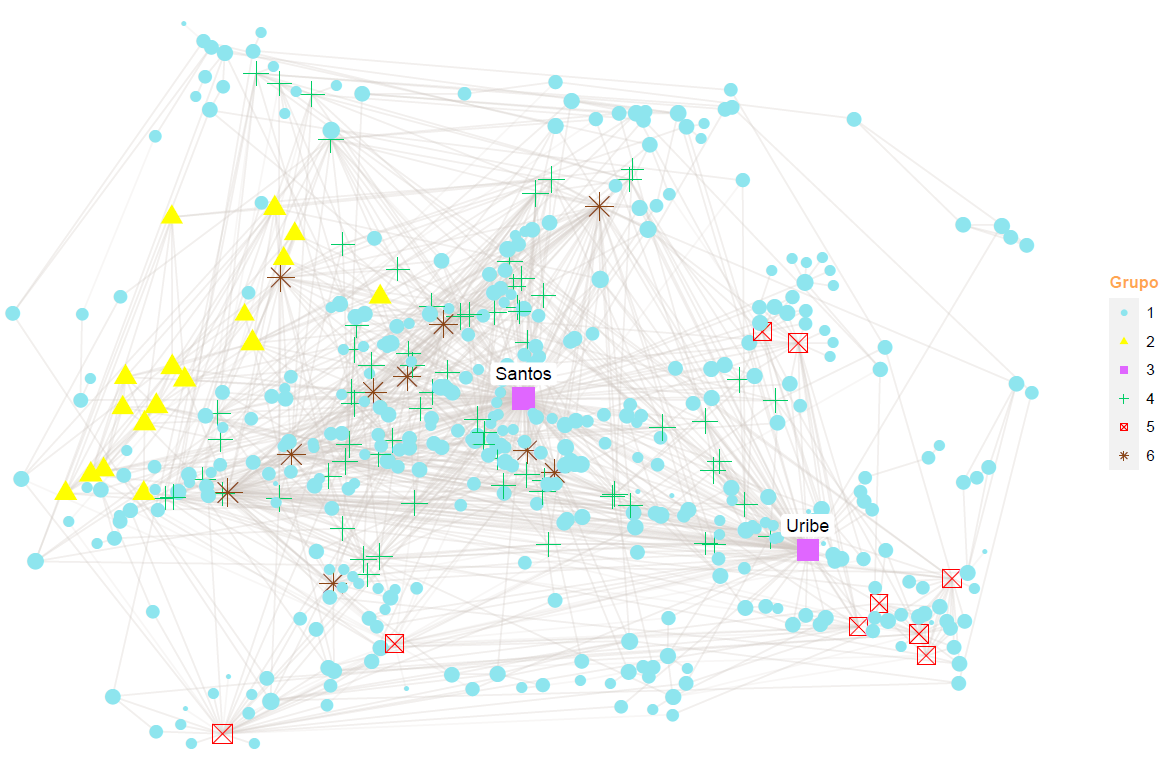}
        }
    \subfigure[Red de alianzas]{
    \centering
        \includegraphics[scale=0.64]{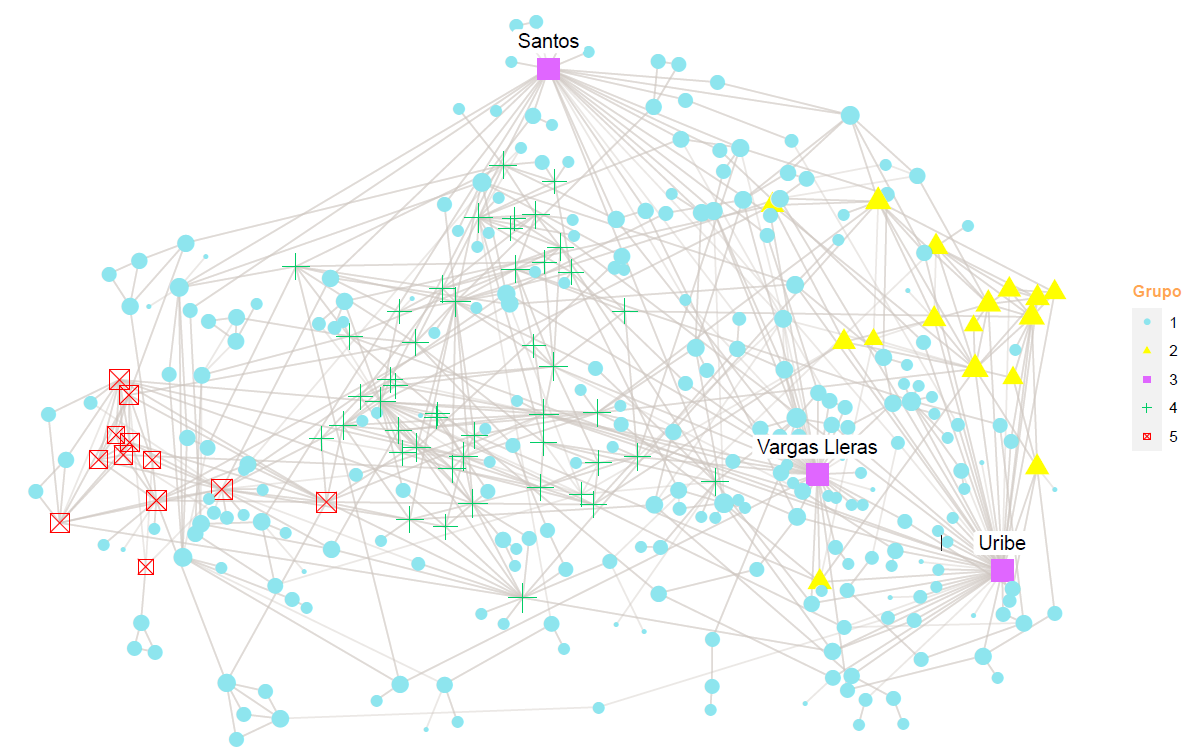}
        }
    \caption{\textit{Partición inducida por el modelo de bloques estocásticos para la red de trabajo y de alianzas.}}
    \label{fig:particiones}
\end{figure}

\begin{figure}[H]
\centering
    \subfigure[Red de trabajo]{
    \centering
        \includegraphics[scale=0.45]{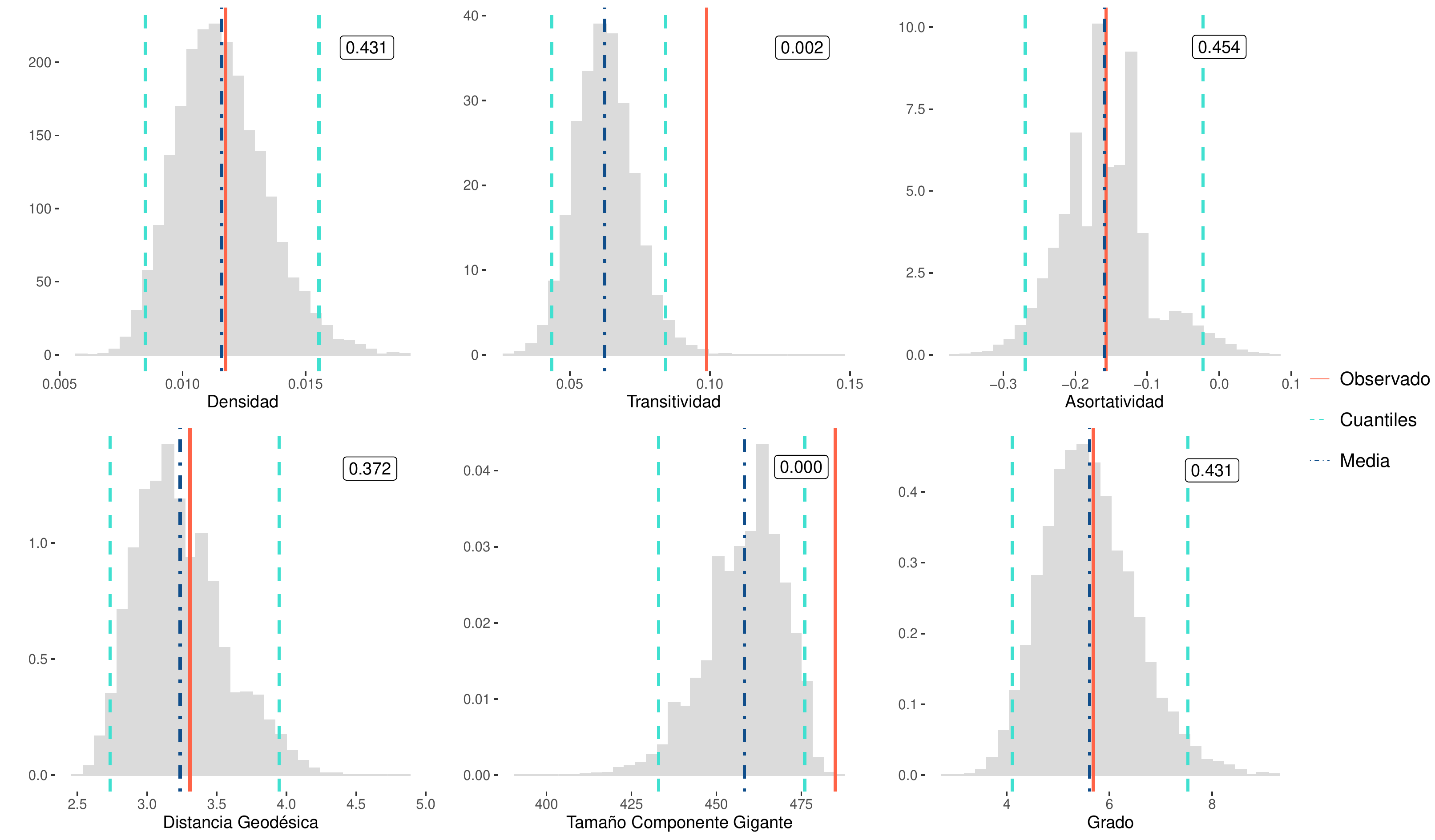}
        }
    \subfigure[Red de alianzas]{
    \centering
        \includegraphics[scale=0.45]{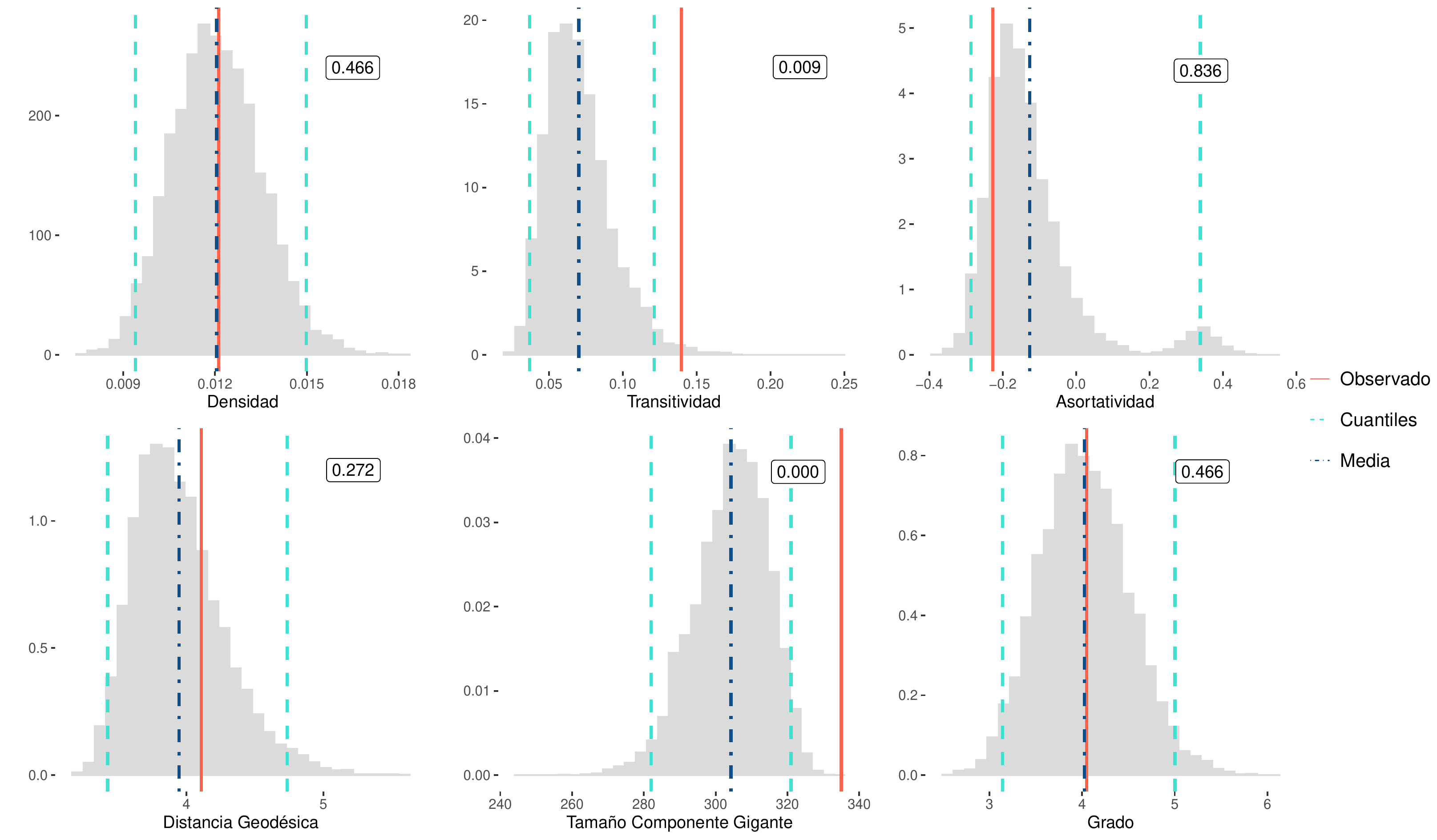}
        }
    \caption{\textit{Distribución predictiva posterior de un conjunto de estadísticos de prueba. En los gráficos la línea roja representa el estadístico calculado a partir de las observaciones, la línea azul oscura la media posterior y las líneas aguamarina los límites del intervalo de credibilidad al 95\% basado en percentiles. El valor numérico corresponde al valor $p$ predictivo posterior.}}
    \label{fig:bondad_ajuste}
\end{figure}

Por último, verificamos la bondad de ajuste del modelo a través de simulaciones que recrean la proporción de los grupos y las probabilidades de interacción obtenidas con el mismo. Las estadísticas
descriptivas de las redes simuladas contrastadas con las de las redes originales se presentan en la Figura \ref{fig:bondad_ajuste}. Evidenciamos para ambas redes que el modelo captura correctamente la densidad, la asortatividad, la distancia geodésica promedio y el grado medio de las mismas. Para los estadísticos mencionados el valor p predictivo posterior (ppp) no asume
valores extremos cercanos a 0 o 1, lo cual indica que el modelo recupera tales características de los datos \cite[ver][para más detalles sobre la bondad de ajuste de modelos Bayesianos]{luquesosa2023}. Sin embargo, el modelo tiende a subestimar ligeramente la conectividad de las redes porque los valores predichos de la transitividad y la cantidad de nodos efectivamente conectados (tamaño de la componente gigante) son algo menores que los valores observados. No obstante, lo anterior no interfiere con la conformación de las comunidades presentadas anteriormente.

\section{Discusión} \label{Sec4}

Implementamos métodos de redes desde dos perspectivas, palabras y actores políticos. Desde la primera perspectiva, identificamos términos frecuentes e importantes de cada una de las alocuciones presidenciales. Por ejemplo, evidenciamos en el discurso de los cuatro mandatarios términos de carácter social que exaltan la imperiosa necesidad de combatir la violencia, el conflicto armado, la pobreza y desigualdad en el país. A diferencia de sus antecesores, Petro utiliza morfemas de género,  términos económicos y ambientales como aspecto distintivo en su acción comunicativa inaugural. Con relación a la polaridad de las palabras, reconocemos que todos los mandatarios utilizan sentimientos antagónicos para trasmitir su mensaje al pueblo. No obstante, observamos un vocabulario más amplio de palabras con polaridad positiva en el discurso de Duque y Petro. En este sentido, motivamos al lector a incorporar otras técnicas de análisis de sentimientos para ahondar en la intención persuasiva de los gobernantes \cite[e.g.,][]{mutinda2023sentiment} y ampliar la discusión sobre la distinción entre sentimiento y postura política del individuo \cite[e.g.,][]{bestvater2023sentiment}.

En la misma línea, examinamos redes de bigramas y skip-gramas e identificamos temáticas centrales en cada discurso. Específicamente, reconocemos en el discurso de Uribe y Santos un uso recurrente de pares de términos característicos de su campaña presidencial. Por ejemplo, seguridad democrática y unidad nacional. Por su parte, Duque utiliza combinaciones de palabras que enfatizan aspectos sociales y de justicia, mientras que Petro acentúa en pares de términos que aluden a sostenibilidad, economía y ambiente. La medida de centralidad propia de los nodos que componen las redes construidas a partir de bigramas y skip-gramas, exhibe para el caso de Santos, Duque, y Petro, a Colombia como la palabra de mayor importancia en su alocución. Lo anterior indica apelación a un discurso patriótico para obtener respaldo social y político a través del refuerzo de la identidad colectiva \citep{gonzalez2016discurso}.

Adicionalmente, examinamos cohesión en los discursos a través de estadísticas de redes. Postulamos la transitividad como una medida útil para examinar la perdida de estructura gramatical cuando se seleccionan solo algunos términos para representar el documento original \cite[ver][]{rudkowsky2018more}. A través de la medida de densidad argumentamos que los discursos de los cuatro mandatarios comprenden un número amplio de tópicos sin una acentuación puntual en ninguno de ellos. No obstante, es de trabajos futuros implementar modelos de tópicos para explorar los temas que subyacen en las intervenciones de los jefes de Estado \cite[e.g.,][]{grimmer2010bayesian}.

Finalmente, desde la segunda perspectiva, distinguimos a Uribe y Santos como actores notables en la consolidación de relaciones de trabajo y alianzas. En particular, Uribe representa el nodo más importante de la red de alianzas y Santos el vértice predominante de la red de trabajo. Por otro lado, identificamos comunidades de actores políticos y sus respectivas probabilidades de interacción dentro y entre agrupaciones a través de la implementación del modelo de bloques estocásticos. Las particiones de la red de trabajo revelan una alta probabilidad de interacción entre Uribe y Santos. Luego, el modelo recuperan el vínculo laboral de estos dos actores. Por otra parte, la partición de la red de alianzas exalta una alta probabilidad de interacción entre miembros del partido político Polo Democrático Alternativo. Nuestra propuesta permite establecer grupos y explicar la forma cómo los actores
establecen relaciones entre sí. Sin embargo, no examinamos  variaciones de estas agrupaciones a lo largo del tiempo. Por ello, es de investigaciones posteriores analizar la evolución de las comunidades políticas propuestas \cite[e.g.,][]{daniel2023bayesian}.

\newpage
\bibliography{ref}

\begin{thebibliography}{}

\bibitem[Abbe, 2017]{abbe2017community}
Abbe, E. (2017).
\newblock Community detection and stochastic block models: recent developments.
\newblock {\em The Journal of Machine Learning Research}, 18(1):6446--6531.

\bibitem[Alschner, 2023]{alschner2023network}
Alschner, W. (2023).
\newblock Network analysis for the comparative study of judicial behaviour.
\newblock {\em Epstein et al., Oxford Handbook of Comparative Judicial
  Behaviour, Forthcoming}.

\bibitem[Ayeomoni and Akinkuolere, 2012]{ayeomoni2012pragmatic}
Ayeomoni, O.~M. and Akinkuolere, O.~S. (2012).
\newblock A pragmatic analysis of victory and inaugural speeches of president
  umaru musa yar'adua.
\newblock {\em Theory \& Practice in Language Studies}, 2(3).

\bibitem[Bai et~al., 2023]{bai2023web}
Bai, Y., Jia, R., and Yang, J. (2023).
\newblock Web of power: How elite networks shaped war and politics in china.
\newblock {\em The Quarterly Journal of Economics}, 138(2):1067--1108.

\bibitem[Bestvater and Monroe, 2023]{bestvater2023sentiment}
Bestvater, S.~E. and Monroe, B.~L. (2023).
\newblock Sentiment is not stance: target-aware opinion classification for
  political text analysis.
\newblock {\em Political Analysis}, 31(2):235--256.

\bibitem[Blair and Potter, 2023]{blair2023strategic}
Blair, C.~W. and Potter, P.~B. (2023).
\newblock The strategic logic of large militant alliance networks.
\newblock {\em Journal of Global Security Studies}, 8(1):ogac035.

\bibitem[Blei et~al., 2003]{blei2003latent}
Blei, D.~M., Ng, A.~Y., and Jordan, M.~I. (2003).
\newblock Latent dirichlet allocation.
\newblock {\em Journal of machine Learning research}, 3(Jan):993--1022.

\bibitem[Bol{\'\i}var, 2017]{bolivar2017political}
Bol{\'\i}var, A. (2017).
\newblock {\em Political discourse as dialogue: A Latin American perspective}.
\newblock Routledge.

\bibitem[Brett, 2018]{brett2018role}
Brett, R. (2018).
\newblock The role of the victims’ delegations in the santos-farc peace
  talks.
\newblock {\em The Politics of Victimhood in Post-Conflict Societies:
  Comparative and Analytical Perspectives}, pages 267--299.

\bibitem[Bustamante, 2020]{bustamante2020practicas}
Bustamante, L.~M. (2020).
\newblock Pr{\'a}cticas de pol{\'\i}tica exterior colombiana: La
  construcci{\'o}n y el manejo de la pol{\'\i}tica antinarc{\'o}ticos en los
  gobiernos de {\'a}lvaro uribe (2002-2006) y juan manuel santos (2014-2018).
\newblock {\em Perspectivas Internacionales}, 14(1):7--42.

\bibitem[C{\'a}rdenas, 2015]{cardenas2015latin}
C{\'a}rdenas, J. (2015).
\newblock Are latin america's corporate elites transnationally interconnected?
  a network analysis of interlocking directorates.
\newblock {\em Global networks}, 15(4):424--445.

\bibitem[C{\'a}rdenas-T{\'a}mara, 2012]{cardenas2012aparato}
C{\'a}rdenas-T{\'a}mara, F. (2012).
\newblock Aparato discursivo del expresidente {\'a}lvaro uribe v{\'e}lez.
  horizontes mim{\'e}ticos del pensamiento hegem{\'o}nico neoliberal en
  colombia (2002-2010).
\newblock {\em An{\'a}lisis pol{\'\i}tico}, 25(76):139--157.

\bibitem[Daniel~Loyal and Chen, 2023]{daniel2023bayesian}
Daniel~Loyal, J. and Chen, Y. (2023).
\newblock A bayesian nonparametric latent space approach to modeling evolving
  communities in dynamic networks.
\newblock {\em Bayesian Analysis}, 18(1):49--77.

\bibitem[de~Oliveira~Paes, 2023]{de2023networked}
de~Oliveira~Paes, L. (2023).
\newblock Networked territoriality: A processual--relational view on the making
  (and makings) of regions in world politics.
\newblock {\em Review of International Studies}, 49(1):53--82.

\bibitem[Eder, 2023]{eder2023discourse}
Eder, F. (2023).
\newblock Discourse network analysis.
\newblock In {\em Routledge Handbook of Foreign Policy Analysis Methods}.
  Taylor \& Francis.

\bibitem[Espinel and Rodr{\'\i}guez, 2019]{espinel2019polarizacion}
Espinel, {\'O}. A.~P. and Rodr{\'\i}guez, L. M.~R. (2019).
\newblock Polarizaci{\'o}n y demonizaci{\'o}n en la campa{\~n}a presidencial de
  colombia de 2018: an{\'a}lisis del comportamiento comunicacional en el
  twitter de gustavo petro e iv{\'a}n duque.
\newblock {\em Revista humanidades}, 9(1).

\bibitem[Fierro, 2014]{fierro2014alvaro}
Fierro, M.~I. (2014).
\newblock {\'A}lvaro uribe v{\'e}lez populismo y neopopulismo.
\newblock {\em An{\'a}lisis pol{\'\i}tico}, 27(81):127--147.

\bibitem[Garcia-Arteaga and Pellegrino, 2021]{garcia2021network}
Garcia-Arteaga, J.~D. and Pellegrino, V. (2021).
\newblock A network based approach to characterize twenty-first-century
  populism in colombia.
\newblock {\em arXiv preprint arXiv:2102.03429}.

\bibitem[Garifullina et~al., 2021]{garifullina2021inaugural}
Garifullina, D.~B., Khismatullina, L.~G., Giniyatullina, A.~Y., Garaeva, M.~R.,
  and Gimadeeva, A.~A. (2021).
\newblock Inaugural speech as a tool of forming speech portrait of the
  president.
\newblock {\em Linguistics and Culture Review}, 5(S1):413--421.

\bibitem[Gonz{\'a}lez~Salinas, 2016]{gonzalez2016discurso}
Gonz{\'a}lez~Salinas, O.~F. (2016).
\newblock El discurso patri{\'o}tico y el aparato propagand{\'\i}stico que
  sustentaron a la expropiaci{\'o}n petrolera durante el cardenismo.
\newblock {\em Estudios de historia moderna y contempor{\'a}nea de M{\'e}xico},
  pages 88--107.

\bibitem[Grimmer, 2010]{grimmer2010bayesian}
Grimmer, J. (2010).
\newblock A bayesian hierarchical topic model for political texts: Measuring
  expressed agendas in senate press releases.
\newblock {\em Political Analysis}, 18(1):1--35.

\bibitem[Guerra-Molina and Badillo-Sarmiento, 2021]{guerra2021desecuritizacion}
Guerra-Molina, R. and Badillo-Sarmiento, R. (2021).
\newblock Desecuritizaci{\'o}n y securitizaci{\'o}n del narcotr{\'a}fico en el
  marco del acuerdo de paz en colombia.
\newblock {\em URVIO Revista Latinoamericana de Estudios de Seguridad}, pages
  8--27.

\bibitem[Haim, 2016]{haim2016alliance}
Haim, D.~A. (2016).
\newblock Alliance networks and trade: The effect of indirect political
  alliances on bilateral trade flows.
\newblock {\em Journal of Peace Research}, 53(3):472--490.

\bibitem[Haselmayer and Jenny, 2017]{haselmayer2017sentiment}
Haselmayer, M. and Jenny, M. (2017).
\newblock Sentiment analysis of political communication: Combining a dictionary
  approach with crowdcoding.
\newblock {\em Quality \& Quantity}, 51:2623--2646.

\bibitem[Jockers, 2017]{jockers2017package}
Jockers, M. (2017).
\newblock Package ‘syuzhet’.
\newblock {\em URL: https://cran. r-project. org/web/packages/syuzhet}.

\bibitem[Khoo and Johnkhan, 2018]{khoo2018lexicon}
Khoo, C.~S. and Johnkhan, S.~B. (2018).
\newblock Lexicon-based sentiment analysis: Comparative evaluation of six
  sentiment lexicons.
\newblock {\em Journal of Information Science}, 44(4):491--511.

\bibitem[Knoke, 1990]{knoke1990political}
Knoke, D. (1990).
\newblock {\em Political networks: the structural perspective}, volume~4.
\newblock cambridge university press.

\bibitem[Kolaczyk and Cs{\'a}rdi, 2020]{kolaczyk}
Kolaczyk, E.~D. and Cs{\'a}rdi, G. (2020).
\newblock {\em Statistical analysis of network data with R}, volume~65.
\newblock Springer.

\bibitem[Latouche et~al., 2011]{latouche2011overlapping}
Latouche, P., Birmel{\'e}, E., and Ambroise, C. (2011).
\newblock Overlapping stochastic block models with application to the french
  political blogosphere.
\newblock {\em The Annals of Applied Statistics}, 5(1):309--336.

\bibitem[Leifeld, 2017]{leifeld2017discourse}
Leifeld, P. (2017).
\newblock Discourse network analysis.
\newblock {\em The Oxford handbook of political networks}, pages 301--326.

\bibitem[Leifeld, 2020]{leifeld2020policy}
Leifeld, P. (2020).
\newblock Policy debates and discourse network analysis: A research agenda.
\newblock {\em Politics and Governance}, 8(2):180--183.

\bibitem[Luque and Sosa, 2023]{luquesosa2023}
Luque, C. and Sosa, J. (2023).
\newblock Bayesian analysis for social science research.
\newblock {\em arXiv preprint arXiv:2306.11966}.

\bibitem[Mutinda et~al., 2023]{mutinda2023sentiment}
Mutinda, J., Mwangi, W., and Okeyo, G. (2023).
\newblock Sentiment analysis of text reviews using lexicon-enhanced bert
  embedding (lebert) model with convolutional neural network.
\newblock {\em Applied Sciences}, 13(3):1445.

\bibitem[Nabuco-Martuscelli and Duarte-Villa, 2018]{Martuscelli2018}
Nabuco-Martuscelli, P. and Duarte-Villa, R. (2018).
\newblock Child soldiers as peace-builders in colombian peace talks between the
  government and the {FARC}{\textendash}{EP}.
\newblock {\em Conflict, Security {\&} Development}, 18(5):387--408.

\bibitem[Nasi, 2010]{nasi2010saboteadores}
Nasi, C. (2010).
\newblock Saboteadores de los procesos de paz en colombia.
\newblock {\em Conflicto armado, seguridad y construcci{\'o}n de paz en
  Colombia}, pages 97--129.

\bibitem[Nowicki and Snijders, 2001]{nowicki2001estimation}
Nowicki, K. and Snijders, T. A.~B. (2001).
\newblock Estimation and prediction for stochastic blockstructures.
\newblock {\em Journal of the American statistical association},
  96(455):1077--1087.

\bibitem[Oladayo, 2023]{oladayo2023language}
Oladayo, M.~M. (2023).
\newblock Language and democracy in nigeria: A stylistic study of late
  president umar musa yar’adua’s inaugural speech of may 29, 2007.
\newblock {\em Journal of Linguistics and Communication Studies}, 2(1):15--21.

\bibitem[Oner and Shehadeh, 2023]{oner2023populist}
Oner, I. and Shehadeh, L. (2023).
\newblock Populist discourse beyond the borders: The case of erdogan and
  chavez.
\newblock {\em Populism}, 1(aop):1--27.

\bibitem[Porter, 2006]{porter2006algorithm}
Porter, M.~F. (2006).
\newblock An algorithm for suffix stripping.
\newblock {\em Program}, 40(3):211--218.

\bibitem[Qader et~al., 2019]{qader2019overview}
Qader, W.~A., Ameen, M.~M., and Ahmed, B.~I. (2019).
\newblock An overview of bag of words; importance, implementation,
  applications, and challenges.
\newblock In {\em 2019 international engineering conference (IEC)}, pages
  200--204. IEEE.

\bibitem[Rice and Zorn, 2021]{rice2021corpus}
Rice, D.~R. and Zorn, C. (2021).
\newblock Corpus-based dictionaries for sentiment analysis of specialized
  vocabularies.
\newblock {\em Political Science Research and Methods}, 9(1):20--35.

\bibitem[R{\'\i}os and Morales, 2022]{rios2022discurso}
R{\'\i}os, J. and Morales, J. (2022).
\newblock El discurso de iv{\'a}n duque sobre el acuerdo con las farc-ep en el
  escenario internacional.
\newblock {\em Revista Opera}, pages 123--142.

\bibitem[Rudkowsky et~al., 2018]{rudkowsky2018more}
Rudkowsky, E., Haselmayer, M., Wastian, M., Jenny, M., Emrich, {\v{S}}., and
  Sedlmair, M. (2018).
\newblock More than bags of words: Sentiment analysis with word embeddings.
\newblock {\em Communication Methods and Measures}, 12(2-3):140--157.

\bibitem[Sutterl{\"u}tti and Meretz, 2023]{sutterlutti2023reform}
Sutterl{\"u}tti, S. and Meretz, S. (2023).
\newblock Reform and revolution.
\newblock In {\em Make Capitalism History: A Practical Framework for Utopia and
  the Transformation of Society}, pages 41--72. Springer.

\bibitem[Victor et~al., 2017]{victor2016introduction}
Victor, J.~N., Montgomery, A.~H., and Lubell, M. (2017).
\newblock {\em The Oxford handbook of political networks}.
\newblock Oxford University Press.

\bibitem[Villarraga, 2012]{villarraga2012analisis}
Villarraga, L.~Y. (2012).
\newblock An{\'a}lisis del discurso de posesi{\'o}n de juan manuel santos: la
  ideolog{\'\i}a de la unidad nacional.
\newblock {\em Forma y Funci{\'o}n}, 25(1):35--51.

\bibitem[Ward et~al., 2011]{ward2011network}
Ward, M.~D., Stovel, K., and Sacks, A. (2011).
\newblock Network analysis and political science.
\newblock {\em Annual Review of Political Science}, 14:245--264.

\bibitem[Yun et~al., 2014]{yun2014climate}
Yun, S.-J., Ku, D., and Han, J.-Y. (2014).
\newblock Climate policy networks in south korea: alliances and conflicts.
\newblock {\em Climate policy}, 14(2):283--301.

\bibitem[Zech and Gabbay, 2016]{zech2016social}
Zech, S.~T. and Gabbay, M. (2016).
\newblock Social network analysis in the study of terrorism and insurgency:
  From organization to politics.
\newblock {\em International Studies Review}, 18(2):214--243.

\bibitem[Zhang et~al., 2023]{zhang2023deep}
Zhang, X., Goel, R.~K., Jiang, J., and Capasso, S. (2023).
\newblock Do deep regional trade agreements strengthen anti-corruption? a
  social network analysis.
\newblock {\em The World Economy}.

\end{thebibliography}
\bibliographystyle{apalike}

\end{document}